\documentstyle[12pt]{article}
\begin{document}
\begin{center} {\large\bf EFFECT OF BINDING IN DEEP INELASTIC
SCATTERING REVISITED} \end{center}
\vskip 1em
\begin{center} {\large Felix M. Lev} \end{center}
\vskip 1em
\begin{center} {\it Laboratory of Nuclear
Problems, Joint Institute for Nuclear Research, Dubna, Moscow region
141980 Russia (E-mail:  lev@nusun.jinr.dubna.su)} \end{center}
\vskip 1em
\begin{abstract}
 In the Bjorken limit of the present theory of deep inelastic
scattering (DIS) the structure functions (up to anomalous dimensions
and perturbative QCD corrections) are described by the parton model.
However the current operator in the parton model does not properly
commute with the representation operators corresponding to the Lorentz
group, space reflection and time reversal. To investigate the violation
of these symmetries in the parton model we consider a model in which
the current operator explicitly satisfies extended Poincare invariance
and current conservation. It is shown that due to binding of quarks in
the nucleon the Bjorken variable $x$ no longer can be interpreted as the
internal light cone momentum fraction $\xi$ even in the Bjorken limit. As
a result, the data on DIS alone do not make it possible to determine the
$\xi$ distribution of quarks in the nucleon. We also consider a qualitative
explanation of the fact that in the parton model the values given by the
sum rules exceed the corresponding experimental quantities while the
quark contribution to the nucleon momentum and spin is underestimated.
\vskip 0.3em

\end{abstract}
\vskip 1em

\section{The statement of the problem}
\label{S1}

 A full description of any relativistic quantum system implies in
particular that one can construct the momentum and angular momentum
operators $({\hat P}^{\mu},{\hat M}^{\mu\nu})$ $(\mu,\nu=0,1,2,3$,
${\hat M}^{\nu\mu}=-{\hat M}^{\mu\nu})$ which are the generators of
the (pseudo)unitary representation of the Poincare group for this
system. We shall always assume that the commutation relations between
the generators are realized in the form
$$[{\hat P}^{\mu},{\hat P}^{\nu}]=0, \quad [{\hat M}^{\mu\nu},
{\hat P}^{\rho}]= -\imath({\eta}^{\mu\rho}{\hat P}^{\nu}-
{\eta}^{\nu\rho}{\hat P}^{\mu}),$$
\begin{equation}
[{\hat M}^{\mu\nu},{\hat M}^{\rho\sigma}]=-\imath ({\eta}^{\mu\rho}
{\hat M}^{\nu\sigma}+{\eta}^{\nu \sigma}{\hat M}^{\mu\rho}-{\eta}^
{\mu\sigma}{\hat M}^{\nu\rho}-{\eta}^{\nu\rho}{\hat M}^{\mu\sigma})
\label{1}
\end{equation}
where the metric tensor in Minkowski space has the nonzero components
$\eta^{00}=-\eta^{11}=-\eta^{22}=-\eta^{33}=1$ and we use the system of
units with $\hbar=c=1$. We use $(P^{\mu},M^{\mu\nu})$ to denote the
corresponding generators in the case when all interactions in the
system under consideration are turned off.

 In the relativistic theory, in contrast with the nonrelativistic one,
it is not possible to realize the operators
$({\hat P}^{\mu},{\hat M}^{\mu\nu})$ in such a way that only the
Hamiltonian ${\hat P}^0$ is interaction dependent while all the other
nine generators are free. Indeed, suppose that ${\hat P}^0$ is interaction
dependent and consider the relation
$[{\hat M}^{0i},{\hat P}^k]=-\imath \delta_{ik}{\hat P}^0$ $(i,k=1,2,3)$
which follows from Eq. (\ref{1}). Then it is obvious that if all the
operators ${\hat P}^k$ are free then all the operators ${\hat M}^{0i}$ are
inevitably interaction dependent
and {\it vice versa}, if all the operators ${\hat M}^{0i}$ are free then
all the operators ${\hat P}^k$ are inevitably interaction dependent.
According to the Dirac classification \cite{Dir}, the realization of the
algebra (\ref{1}) in such a way that the operators
$({\hat P}^0,{\hat M}^{0i})$  are interaction
dependent and the other six generators of the Poincare group are free
is called the instant form of dynamics, while the point form implies that
all the components ${\hat P}^{\mu}$ are interaction dependent and
${\hat M}^{\mu\nu}=M^{\mu\nu}$. Instead of the $0,z$ components of
four-vectors we also can work with the $\pm$ components defined as
$p^{\pm}=(p^0\pm p^z)/\sqrt{2}$. Suppose that ${\hat P}^-$ is interaction
dependent and the other components of ${\hat P}^{\mu}$ are free. Then, as
follows from Eq. (\ref{1}), $[{\hat M}^{-j},{\hat P}^l]=-\imath
\delta_{jl} {\hat P}^-$ ($j=1,2$) and hence all the operators
${\hat M}^{-j}$ are inevitably interaction dependent. The realization of
the algebra (\ref{1}) in such a way that $({\hat P}^-,{\hat M}^{-j})$
are interaction dependent and the other seven generators are free
is called the front form \cite{Dir}.  Of course, physical results
should not depend on the choice of the form of dynamics and in the
general case all the ten generators can be interaction dependent.

 We can also consider extended relativistic invariance
which implies that the generators properly commute not only with each
other but also with the representation operators ${\hat U}_P$ and
${\hat U}_T$ corresponding to space reflection and time reversal. A
possible choice in the instant and point forms is ${\hat U}_P=U_P$ and
${\hat U}_T=U_T$, but in the front form the operators ${\hat U}_P$ and
${\hat U}_T$ should be necessarily interaction
dependent. This follows in particular from the relations
\begin{equation}
{\hat U}_PP^+{\hat U}_P^{-1}=
{\hat U}_TP^+{\hat U}_T^{-1}={\hat P}^-
\label{2}
\end{equation}
As noted by Coester \cite{Coes}, the interaction dependence of the
operators ${\hat U}_P$ and ${\hat U}_T$ in the front form does not mean
that there are discrete dynamical symmetries in addition to the
rotations about transverse axes.
Indeed, the discrete transformation $P_2$ such that
$P_2\, x:= \{x^0,x_1,-x_2,x_3\}$ leaves the light front $x^+=0$ invariant
and therefore $P_2$ can be chosen free.
The full space reflection $P$ is the product of $P_2$ and a rotation about
the 2-axis by $\pi$. Thus it is not an independent dynamical transformation
in addition to the rotations about transverse axes.
Similarly the transformation $TP$ leaves $x^+=0$ invariant and
$T=(TP)P_2R_2(\pi)$.

 To describe some process of deep inelastic scattering
(DIS) it is necessary to construct the operator of the electromagnetic or
weak current responsible for the transition $nucleon\rightarrow hadrons$
in this process. Let ${\hat J}^{\mu}(x)$ be such an operator where $x$ is
a point in Minkowski space. Translational invariance of the current
operator implies that
\begin{equation}
exp(-\imath {\hat P}_{\mu}a^{\mu}){\hat  J}^{\mu}(x)
exp(\imath {\hat P}_{\mu}a^{\mu})={\hat J}^{\mu}(x-a)
\label{3}
\end{equation}
and Lorentz invariance implies that
\begin{equation}
[{\hat M}^{\mu\nu},
{\hat J}^{\rho}(x)]= -\imath\{(x^{\mu}\partial^{\nu}-x^{\nu}\partial^{\mu})
{\hat J}^{\rho}(x)+\eta^{\mu\rho}{\hat J}^{\nu}(x)-\eta^{\nu\rho}
{\hat J}^{\mu}(x)\}
\label{4}
\end{equation}
The electromagnetic current operator should also satisfy the continuity
equation $\partial {\hat J}^{\mu}(x)/\partial x^{\mu}=0$. As follows from
Eq. (\ref{3}), this equation can be written in the form
\begin{equation}
[{\hat  J}^{\mu}(x),{\hat P}_{\mu}]=0
\label{5}
\end{equation}

 Finally, the usual requirement is that the theory should be local in
the sense that the commutator $[{\hat J}^{\mu}(x),{\hat J}^{\nu}(y)]$
should necessarily vanish when $x-y$ is a space-like vector.
Let us note however that if a theory is nonlocal in the above sense,
this does not necessarily imply that it is nonphysical. Indeed, as it
has become clear already in 30th, in relativistic quantum theory there
is no operator possessing all the properties of the position operator.
In particular, the quantity $x$ in the Lagrangian density $L(x)$ is
not the coordinate, but some parameter which becomes the coordinate only
in the classical limit. Therefore the physical condition is that
only the macroscopic locality should be satisfied, i.e. the
above commutator should vanish when $|(x-y)^2|\rightarrow \infty$ but
$(x-y)^2<0$.

 In the framework of the scattering theory it is sufficient to ensure
relativistic invariance of the S-matrix and hence it is sufficient to
consider Eqs. (\ref{1}-\ref{5}) only in the scattering space of the
system under consideration. In QED the electrons, positrons and photons
are the fundamental particles, and the scattering
space is the space of these almost free particles ("in" or "out" space).
Therefore it is sufficient to deal only with ${\hat P}_{ex}^{\mu},
{\hat M}_{ex}^{\mu\nu}$ where "ex" stands either for "in" or "out".
However in QCD the scattering space by no means can be considered as a
space of almost free fundamental particles --- quarks and gluons. For
example, even if the scattering space consists of one particle (say the
nucleon), this particle is the bound state of quarks and gluons, and the
operators ${\hat P}^{\mu}$ and ${\hat M}^{\mu\nu}$ considerably differ
from $P^{\mu}$ and $M^{\mu\nu}$. It is well-known that perturbation
theory does not apply to bound states and therefore
${\hat P}^{\mu}$ and ${\hat M}^{\mu\nu}$ cannot be determined in the
framework of perturbation theory. For these reasons we will be interested
in cases when the representation operators correspond to the full
generators ${\hat P}^{\mu}$ and ${\hat M}^{\mu\nu}$.

 In quantum field theory the only known way of constructing the
operators ${\hat P}^{\mu}$, ${\hat M}^{\mu\nu}$ and ${\hat J}^{\mu}(x)$
is canonical formalism. However this formalism ignores the fact that
local field operators are operator valued distributions and therefore
products of these operators at coinciding points are not well defined.
Moreover, canonical formalism contradicts the Haag theorem \cite{Haag}.
A detailed discussion of these problems can be found, for example, in
the well-known monographs \cite{BLOT}. It has been also shown in a
vast literature (see e.g. Refs. \cite{Schw,AD,JP,J,MW}) that the
canonical treatment of equal time commutation relations between the
current operators can often lead to incorrect results. The results of
the above references show that it is often premature to trust assumptions
which may seem physical or natural.

 In Sect. \ref{S2} we argue that the assumptions used in the present
theory of DIS are not substantiated and the current operator contains
a nontrivial nonperturbative part which contributes to DIS even in the
Bjorken limit. Since this contribution cannot be determined at the
present stage of QCD, it is reasonable to consider models in which the
operators ${\hat P}^{\mu}$, ${\hat M}^{\mu\nu}$ and ${\hat J}^{\mu}(x)$
are well defined and can be explicitly constructed but of course one
should understand that such models cannot have any pretensions to
be fundamental. In Sect. \ref{S3} the explicit construction of the
operators ${\hat P}^{\mu}$ and ${\hat M}^{\mu\nu}$ in the
framework of our models is considered. In Sect. \ref{S4} the
well-known results that the parton model is a consequence of impulse
approximation (IA) for the current operator in the front form are
reproduced but in our consideration it is clear that such a current
operator does not properly commute with the Lorentz group generators
and the operators corresponding to the discrete symmetries. As shown
in Sect. \ref{S5}, the idea that at large momentum transfer the
current operator can be taken in IA can be consistently realized in
the point form and the corresponding results considerably differ
from the results of the parton model. In Sect. \ref{S6} the
difference between the parton model sum rules and the sum rules in
our model is considered and Sect. \ref{S7} is discussion.

\section{Problems in the present theory of DIS}
\label{S2}

 In addition to the examples considered in Refs. \cite{AD,JP,J,MW}
we now give one more example which shows that the reader thinking
that it is not reasonable to worry about mathematical rigor will
be confronted with the following contradiction.

 In QED and QCD the electromagnetic current operator is usually
written in the form
${\hat J}^{\mu}(x)={\cal N}\{{\hat {\bar \psi}}(x)\gamma^{\mu}
{\hat \psi}(x)\}$
where ${\cal N}$ stands for the normal product, ${\hat \psi}(x)$ is the
Heisenberg operator of the Dirac field and for simplicity we do not
write flavor operators and color and flavor indices. As noted above,
such a definition ignores the fact that ${\hat J}^{\mu}(x)$ should be
an operator valued distribution. Then we can
consider ${\hat J}^{\mu}(x)$ at fixed values of $x$, for example at $x=0$,
and formally it follows from Eq. (\ref{4}) that
\begin{equation}
[{\hat M}^{\mu\nu},{\hat J}^{\rho}(0)]=-\imath
(\eta^{\mu\rho}{\hat J}^{\nu}(0)-\eta^{\nu\rho}{\hat J}^{\mu}(0))
\label{6}
\end{equation}

 Let us now take into account that in the framework of canonical
quantization the Heisenberg and Schrodinger pictures at $x=0$ are the
same and therefore ${\hat J}^{\mu}(0)$ does not depend on the
interaction, i.e. ${\hat J}^{\mu}(0)=J^{\mu}(0)$ (note that the
free operator $J^{\mu}(x)$ can be considered as a usual operator
valued function of $x$).
The operator ${\hat M}^{\mu\nu}$ can be written as
$M^{\mu\nu}+V^{\mu\nu}$. Since $[M^{\mu\nu},J^{\rho}(0)]$ can be
written by analogy with Eq. (\ref{6}), it is obvious that
$[V^{\mu\nu},J^{\rho}(0)]=0$ and in particular $[V^{0i},J^0(0)]=0$.
The canonical form of $V^{0i}$ in QED is \cite{AB}
\begin{equation}
V^{0i}=e\int\nolimits x^i{\bf A}({\bf x}){\bf J}({\bf x})d^3{\bf x}
\label{7}
\end{equation}
where $e$ is the electric charge, ${\bf A}$ is the operator of the
Maxwell field and ${\bf x}$ is the spatial part of the four-vector
$x$ when $x^0=0$. Therefore $[V^{0i},J^0(0)]=0$ if
\begin{equation}
\int\nolimits x^i {\bf A}({\bf x})[{\bf J}({\bf x}),J^0(0)]
d^3{\bf x}=0
\label{8}
\end{equation}
It is well-known that if the standard equal-time commutation relations
are used naively then the commutator in Eq. (8) vanishes and
therefore this equation is satisfied. However when ${\bf x}\rightarrow
0$ this commutator involves the product of four Dirac fields at
${\bf x}=0$. The famous Schwinger result \cite{Schw} is that
\begin{equation}
[J^i({\bf x}),J^0(0)]=C\frac{\partial}{\partial x^i}\delta({\bf x})
\label{9}
\end{equation}
where $C$ is some (infinite) constant. Therefore Eq. (8) is not
satisfied and the current operator ${\hat J}^{\mu}(x)$ constructed in
the framework of canonical formalism does not satisfy Lorentz invariance.
As a result, the operator ${\hat J}^{\mu}(0)$ (if it exists) cannot be
free.

 Although our example is the demonstration of this property in spinor
QED, as noted above, here this example is not very important.
However our example is important for QCD since it shows that the
algebraic reasons based on Eq. (\ref{6}) are more solid than the reasons
based on formal manipulations with local operators. Indeed, if the
operators ${\hat M}^{\mu\nu}$ are interaction dependent then
${\hat J}^{\mu}(0)$ cannot be free simply because there is no reason for
interaction terms in ${\hat M}^{\mu\nu}$ to commute with all the
operators $J^{\mu}(0)$. Therefore in the instant and front
forms ${\hat J}^{\mu}(0)\neq J^{\mu}(0)$. At the same time it is obvious that
${\hat J}^{\mu}(0)$ can be free in the point form.

\begin{sloppypar}
 The deep inelastic cross-section is fully defined by the hadronic tensor
\begin{equation}
W^{\mu\nu}=\frac{1}{4\pi}\int\nolimits e^{\imath qx} \langle P',\chi|
{\hat J}^{\mu}(x){\hat J}^{\nu}(0)|P',\chi\rangle d^4x
\label{10}
\end{equation}
where $|P',\chi\rangle$ is the state of the initial nucleon with the
four-momentum $P'$ and the internal wave function $\chi$, and $q$ is the
4-momentum of the virtual photon (for simplicity we shall speak about
electromagnetic interactions but the same is valid for weak ones).
The state $|P',\chi\rangle$ is the eigenstate of the
operator ${\hat P}$ with the eigenvalue $P'$ and the eigenstate of the
spin operators ${\hat {\bf S}}^2$ and ${\hat S}^z$ which are constructed
from ${\hat M}^{\mu\nu}$. In particular, ${\hat P}^2|P',\chi\rangle =
m^2 |P',\chi\rangle$ where $m$ is the nucleon mass.
\end{sloppypar}

 The structure of the operator ${\hat P}$ in QCD is rather
complicated (see e.g. Refs. \cite{CCJ,Ynd}) but anyway some of the
components of ${\hat P}$ necessarily contain a part which describes the
interaction of quarks and gluons at large distances where the QCD
running coupling constant $\alpha_s$ is large and perturbation theory
does not apply. In view of the last relation, this
part is responsible for binding of quarks and gluons in the nucleon.
We will call this part the nonperturbative one.

 Suppose that the Hamiltonian ${\hat P}^0$ contains the nonperturbative
part. Then by analogy with the consideration in Sect. \ref{S1}, we can
show that in the instant form all the operators ${\hat M}^{0i}$
inevitably depend on the nonperturbative part and in the point form all
the operators ${\hat P}^k$ inevitably depend on that part.
Analogously if the front form Hamiltonian ${\hat P}^-$ contains the
nonperturbative part then all the operators ${\hat M}^{-j}$ inevitably
depend on this part. Therefore, some components of the operator
${\hat J}^{\mu}(0)$ in the instant and front forms inevitably depend on
the nonperturbative part, and if ${\hat J}^{\mu}(0)=J^{\mu}(0)$ in the point
form then, as follows from Eq. (\ref{3}), the operator ${\hat J}^{\mu}(x)$
in that form inevitably depend on the nonperturbative part.
The fact that the same operators $({\hat P}^{\mu},{\hat M}^{\mu\nu})$
describe the transformations of both the operator ${\hat J}^{\mu}(x)$
and the state $|P',\chi\rangle$ guaranties that $W^{\mu\nu}$ has the
correct transformation properties.

 We see that the relation between the current operator and the state of
the initial nucleon is highly nontrivial. Meanwhile in the present theory
they are considered separately. In the framework of the approach to DIS
based on Feynman diagrams the possibility of the separate consideration
follows from the factorization theorem \cite{ER,EFP,CSS,Sterman} which
asserts in
particular that the amplitude of the lepton-parton interaction
entering into diagrams dominating in DIS depends only on the hard part
of this interaction. Moreover, in leading order in $1/Q$, where
$Q=|q^2|^{1/2}$, one obtains the parton model up to anomalous dimensions
and perturbative QCD corrections which depend on $\alpha_s(Q^2)$.
We will not discuss here the assumptions used in this approach but
note that the struck quark is essentially off-shell and its momentum
${\tilde p}$ is usually such that ${\tilde p}^2$ is of order $\Lambda^2$
where $\Lambda$ is the typical hadron mass. Therefore it is far from
being obvious that interactions of the struck quark can be considered
perturbatively. As noted in Ref. \cite{CSS}, "it is fair to say that
a rigorous treatment of factorization has yet to be provided".

For example, in Ref. \cite{EFP} the proof of the factorization theorem
is based on the expansion
of the lepton-quark amplitude near the point ${\tilde p}^2=0$
(assuming that the quarks are massless) and only a finite number of
terms are taken into account. The result is justified a
{\it posteriori} since it is in agreement with the intuition provided
by the parton model. Indeed, in the parton model the quarks are free
and therefore ${\tilde p}^2=0$. However such a proof by no means
excludes a possibility that actually the ${\tilde p}^2$ dependence is
important even in leading order in $1/Q$ and therefore the correct
result can be obtained only if a sum of an infinite number of diagrams
is calculated.

 Let us now discuss the following question. Since the current
operator depends on the nonperturbative part then this operator depends
on the integrals from the quark and gluon field operators over the
region of large distances where $\alpha_s$ is large. Is this property
compatible with locality? In the framework of canonical formalism
compatibility is obvious but, as noted above, the results based on
canonical formalism are not reliable. Therefore it is not clear whether
in QCD it is possible to construct local electromagnetic and weak
current operators beyond perturbation theory. However the usual
motivation of the parton model is that, as a consequence of asymptotic
freedom (which implies that $\alpha_s(Q^2)\rightarrow \infty$ when
$Q^2\rightarrow \infty$), the partons in the infinite momentum frame
(IMF) are almost free and therefore, at least in leading order in $1/Q$,
the nonperturbative part of ${\hat J}^{\mu}(x)$ is not important.

 We will now consider whether this property can be substantiated in the
framework of the operator product expansion (OPE) developed by Wilson
and others \cite{Wil}. In this framework the product of the currents
entering into Eq. (\ref{10}) can be written symbolically as
\begin{equation}
{\hat J}(x){\hat J}(0)
= \sum_{i} C_i(x^2) x_{\mu_1}\cdots x_{\mu_n}
{\hat O}_i^{\mu_1\cdots \mu_n}
\label{11}
\end{equation}
where $C_i(x^2)$ are the $c$-number Wilson coefficients while the
operators ${\hat O}_i^{\mu_1\cdots \mu_n}$ depend only on field
operators and their covariant derivatives at the origin of Minkowski
space and have the same form as in perturbation theory. For example,
the basis for twist two operators contains in particular
\begin{equation}
{\hat O}_V^{\mu}={\cal N} \{{\hat {\bar \psi}}(0)\gamma^{\mu}{\hat
\psi}(0)\} \quad
{\hat O}_A^{\mu}={\cal N} \{{\hat {\bar \psi}}(0)\gamma^{\mu}
\gamma^5{\hat \psi}(0)\}
\label{12}
\end{equation}

\begin{sloppypar}
 It is important to note that the OPE has been proved only in the
framework of perturbation theory and its validity beyond that theory
is problematic (see the discussion in Ref. \cite{Nov} and references
therein). Therefore if we use Eq. (\ref{11}) in DIS we have to
assume that either nonperturbative effects are not important to some
orders in $1/Q$ and then we can use Eq. (\ref{11}) only
to these orders (see e.g. Ref. \cite{Jaffe}) or it is possible to use
Eq. (\ref{11}) beyond perturbation theory. The question also arises whether
Eq. (\ref{11}) is valid in all the forms of dynamics (as it should be if
it is an exact operator equality) or only in some forms.
\end{sloppypar}

 In the point form all the components of ${\hat P}$ depend on the
nonperturbative part and therefore, in view of Eq. (\ref{3}), it is not
clear
why there is no nonperturbative part in the $x$ dependence of the right
hand side of Eq. (\ref{11}), or if (for some reasons) it is possible to
include the nonperturbative part only into the operators ${\hat O}_i$
then why they have the same form as in perturbation theory.

 One might think that in the front form the $C_i(x^2)$ will be the
same as in perturbation theory due to the following reasons. The
value of $q^-$ in DIS is very large and therefore only a small
vicinity of the light cone $x^+=0$ contributes to the integral (\ref{10}).
The only dynamical component of ${\hat P}$ is ${\hat P}^-$ which enters
into Eq. (\ref{11}) only in the combination ${\hat P}^-x^+$. Therefore
the dependence of ${\hat P}^-$ on the nonperturbative part of the
quark-gluon interaction is of no importance.  These considerations
are not convincing since the integrand is a singular function and the
operator ${\hat J}^{\mu}(0)$ in the front form depends
on the nonperturbative part, but nevertheless we assume that
Eq. (\ref{11}) in the front form is valid.

 If we assume as usual that there is no problem with the convergence
of the OPE series then experiment makes it possible to measure
each matrix element $\langle P',\chi|{\hat O}_i^{\mu_1\cdots \mu_n}|
P',\chi\rangle$.
Let us consider, for example, the matrix element
$\langle P',\chi|{\hat O}_V^{\mu}|P',\chi\rangle$. It transforms as a
four-vector if the Lorentz transformations of ${\hat O}_V^{\mu}$ are
described by the
operators ${\hat M}^{\mu\nu}$ describing the transformations of
$|P',\chi\rangle$, or in other words, by analogy with Eq. (\ref{6})
\begin{equation}
[{\hat M}^{\mu\nu},{\hat O}_V^{\rho}]=-\imath
(\eta^{\mu\rho}{\hat O}_V^{\nu}-\eta^{\nu\rho} {\hat O}_V^{\mu})
\label{13}
\end{equation}
It is also clear that Eq. (\ref{13}) follows from Eqs. (\ref{3}),
(\ref{4}), (\ref{10}) and (\ref{11}).
Since the ${\hat M}^{-j}$ in the front form depend on the
nonperturbative part, the above
considerations  make it possible to conclude that at least
some components ${\hat O}_V^{\mu}$, and analogously some components
${\hat O}_i^{\mu_1\cdot \mu_n}$, also depend on the
nonperturbative part. Since Eq. (\ref{13}) does not contain any $x$ or
$q$ dependence, this conclusion has nothing to do with asymptotic
freedom  and is valid even in leading order in $1/Q$
(in contrast with the statement of the factorization theorem
\cite{ER,EFP,CSS,Sterman}).

 Since the operators ${\hat O}_i^{\mu_1...\mu_n}$ depend on the
nonperturbative part, then by analogy
with the above considerations we conclude that the operators in
Eq. (\ref{12}) are ill-defined and the correct expressions for them involve
integrals from the field operators over large distances where the QCD
coupling constant is large. Therefore it is not clear whether
the operators ${\hat O}_i^{\mu_1...\mu_n}$ are local and whether the
Taylor expansion at $x=0$ is correct, but even it is, the expressions for
${\hat O}_i^{\mu_1...\mu_n}$ will depend on higher twist operators which
contribute even in leading order in $1/Q$.

 To understand whether the OPE is valid beyond perturbation theory
several authors (see e.g. Ref. \cite{Nov} and references therein)
investigated some two-dimensional models and came to different
conclusions. We will not discuss the arguments of these authors but
note that the Lie algebra of the Poincare group for 1+1 space-time
is much simpler than for 3+1 one. In particular, the Lorentz group is
one-dimensional and in the front form the operator $M^{+-}$ is free.
Therefore Eqs. (\ref{6}) and (\ref{13}) in the "1+1 front form" do not
make it possible to conclude that the operators ${\hat J}^{\mu}(0)$ and
${\hat O}_V^{\mu}$ necessarily depend on the nonperturbative part. At
the same time the full space reflection $P$ in the 1+1 front form is an
independent dynamical transformation, in contrast with the situation in
the 3+1 front form (see Sect. \ref{S1}).

 Let us now summarize the results of this section. As follows from
the commutation relations between the current operator and the
generators of the Poincare group representation,
the current operator nontrivially depends on the nonperturbative part
of the interaction responsible for binding of quarks and gluons in the
nucleon. Then the problem arises whether it is possible to construct
a local current operator ${\hat J}^{\mu}(x)$ beyond perturbation
theory and whether the nonperturbative part of the interaction
entering into ${\hat J}^{\mu}(x)$ contributes to DIS in leading order
in $1/Q$. Our consideration shows that the dependence of
${\hat J}^{\mu}(x)$ on the nonperturbative part makes
the OPE problematic. Nevertheless we assume that Eq. (\ref{11}) is valid
beyond perturbation theory but no form of the operators
${\hat O}_i^{\mu_1...\mu_n}$ is prescribed. Then we come to conclusion
that the nonperturbative part indeed contributes to DIS even in leading
order in $1/Q$.

Since the role of nonperturbative effects cannot be neglected even at
very large $Q$, the calculation of their contribution to DIS cannot be
carried out in the framework of QCD (at the present stage of this
theory). For this reason the remainder of this paper is devoted to the
investigation of DIS in models where all the operators in question can
be explicitly constructed and it is possible to explicitly verify that
they satisfy Eqs. (\ref{1}-\ref{5}).

\section{Representations of the Poincare group for systems of
interacting particles}
\label{S3}

 As shown by several authors (see e.g. Refs.
\cite{sok,CP,M,lev1,KeiPol,lev2}), the representation generators
satisfying Eqs. (\ref{1}) and (\ref{2}) can be explicitly constructed
in the framework of relativistic quantum mechanics (RQM) of systems with
a fixed number of interacting particles and, as shown in Refs.
\cite{KlPol,lev}, in this framework it is also possible to
explicitly construct the current operator satisfying Eqs.
(\ref{3}-\ref{5}).

 Of course, as already noted, any model consideration cannot have any
pretensions to be fundamental. Nevertheless in this paper we argue that
the consideration in the framework of RQM can shed light on the problem
why the effect of binding in DIS is important even at very large values
of $Q$.

 For the consideration of a system of interacting particles it is useful
to consider first kinematical relations in the corresponding system
of free particles and then consider the introduction of the interaction
into this system.

 Consider a system of $N$ free particles with the 4-momenta $p_i$,
masses $m_i$, spins $s_i$, and the electric charges $e_i$
($i=1,...N$). We also use $\sigma_i$ to denote the projection of the
spin of particle $i$ on the $z$ axis and use $\bot$ to denote the
projection of the three-dimensional vectors onto the plane $xy$.
For definiteness we assume that
$m_i>0$, but this assumption is not crucial. Formally the subsequent
consideration will not depend on whether $N$ is finite or infinite
but in the last case essential efforts should be made to prove that
all the operators in question are well defined. For this reason we
assume that $N$ is final.

 The Hilbert space $H$ for the system under consideration is the space
of functions $\varphi({\bf p}_{1\bot},p_1^+,\sigma_1,...
{\bf p}_{N\bot},p_N^+,\sigma_N)$ such that
\begin{equation}
\sum_{\sigma_1...\sigma_N}\int\nolimits |\varphi({\bf p}_{1\bot},p_1^+,
\sigma_1,...{\bf p}_{N\bot},p_N^+,\sigma_N)|^2 \prod_{i=1}^{N}
d\rho({\bf p}_{i\bot},p_i^+)\quad< \infty
\label{14}
\end{equation}
where
\begin{equation}
d\rho({\bf p}_{\bot},p^+)= \frac{d^2{\bf p}_{\bot}dp^+}
{2(2\pi)^3p^+}
\label{15}
\end{equation}

\begin{sloppypar}
 We define $P=p_1+...+p_N$, $M_0=|P|\equiv (P^2)^{1/2}$, and $G=PM_0^{-1}$.
Let $\beta(G) \equiv \beta({\bf G}_{\bot},G^+)\in SL(2,C)$ be the matrix
with the components
\begin{eqnarray}
&&\beta_{11}=\beta_{22}^{-1}=2^{1/4}(G^+)^{1/2},\quad
\beta_{12}=0,\nonumber\\
&&\beta_{21}=(G^x+\imath G^y)\beta_{22},
\label{16}
\end{eqnarray}
$L(\beta)$ be the Lorentz transformation corresponding to
$\beta\in SL(2,C)$ and
\begin{equation}
k_i=L[\beta(G)]^{-1}p_i \quad (i=1,...N)
\label{17}
\end{equation}
\end{sloppypar}

\begin{sloppypar}
The four-vectors $p_i$ have the components $(\omega_i({\bf p}_i),
{\bf p}_i)$, and the four-vectors $k_i$ have the components
$(\omega_i({\bf k}_i),{\bf k}_i)$ where $\omega_i({\bf k})=(m_i^2+
{\bf k}^2)^{1/2}$. In turn, only $N-1$ vectors ${\bf k}_i$ are
independent since, as follows from Eqs. (\ref{16}) and (\ref{17}),
${\bf k}_1+...+{\bf k}_N=0$. Therefore $L[\beta(G)]$ has the meaning
of the boost, and ${\bf k}_i$ are the momenta in the c.m. frame.
It is easy to show that $M_0=\omega_1({\bf k}_1)+...
+\omega_N({\bf k}_N)$.
\end{sloppypar}

\begin{sloppypar}
 Let us define
\begin{equation}
d\rho^F(int)=2(2\pi)^3 M_0\delta^{(3)}({\bf k}_1+\cdots +{\bf k}_N)
 \prod_{i=1}^{N}d\rho({\bf k}_{i\bot},k_i^+)
\label{18}
\end{equation}
We also define the "internal' space $H_{int}^F$ as the space of
functions $\chi^F({\bf k}_1,\sigma_1,...{\bf k}_N,\sigma_N)$
such that
\begin{equation}
||\chi^F||_F^2=\sum_{\sigma_1...\sigma_N}\int\nolimits
|\chi({\bf k}_1,\sigma_1,...{\bf k}_N,\sigma_N)|^2
d\rho^F(int)\quad < \infty
\label{19}
\end{equation}
Then the space of functions satisfying Eq. (\ref{14}) can be realized
as the space of functions $\varphi ({\bf P}_{\bot},P^+)$ with the range
in $H_{int}^F$ and such that
\begin{equation}
\int\nolimits ||\varphi({\bf P}_{\bot},P^+)||_F^2
d\rho({\bf P}_{\bot},P^+)\quad < \infty
\label{20}
\end{equation}
\end{sloppypar}

 For noninteracting particles the representation generators of the
Poincare group are equal to sums of the corresponding one-particle
generators. A direct calculation of these sums shows that in the
variables ${\bf P}_{\bot},P^+,{\bf k}_1,...{\bf k}_N$ the generators
have the form
\begin{eqnarray}
&&P^+=P^+,\quad {\bf P}_{\bot}={\bf P}_{\bot},\quad P^-=
\frac{M_0^2+{\bf P}_{\bot}^2}{2P^+}, \nonumber\\
&&M^{+-}=\imath P^+\frac{\partial}{\partial P^+},\quad
M^{+j}=-\imath P^+\frac{\partial}{\partial P^j}, \nonumber\\
&& M^{xy}=l^z({\bf P}_{\bot})+S^z, \quad
M^{-j}=-\imath(P^j\frac{\partial}{\partial P^+}+
P^-\frac{\partial}{\partial P^j})- \nonumber\\
&&\frac{\epsilon_{jl}}{P^+}(M_0S^l+P^lS^z)
\label{21}
\end{eqnarray}
Here the first expression implies that the generator $P^+$ is equal to
the operator of multiplication by the variable $P^+$ defined above and
the second expression should be understood analogously. The indices
$j,l$ take the values $1,2$ and $\epsilon_{jl}$ is the antisymmetric
tensor with the components $\epsilon_{12}=-\epsilon_{21}=1$,
$\epsilon_{11}=\epsilon_{22}=0$. We also use
${\bf l}({\bf P})=-\imath {\bf P}\times (\partial / \partial {\bf P})$
to denote the orbital angular-momentum operator and ${\bf S}$ to denote
the $N$-particle spin operator. The latter acts only through the
variables of the space $H_{int}^F$ and the explicit form of ${\bf S}$
is of no importance for us. For $N=2$ and $N=3$ this form and a detailed
derivation of Eq. (\ref{21}) can be found, for example, in Refs.
\cite{Ter,BKT,lev2}; the generalization to the case of arbitrary $N$ is
obvious.

\begin{sloppypar}
 As follows from Eq. (\ref{17})
\begin{equation}
\xi_i\equiv \frac{p_i^+}{P^+}=\frac{\omega_i({\bf k}_i)+k_i^z}
{M_0({\bf k}_1,...{\bf k}_N)}\in (0,1)
\label{22}
\end{equation}
Therefore $({\bf k}_{1\bot},\xi_1,...{\bf k}_{N\bot},\xi_N)$ also is a
possible choice of the internal momentum variables, and these variables
are constrained by the relations ${\bf k}_{1\bot}+...+{\bf k}_{N\bot}=0$,
$\xi_1+...+\xi_N=1$.
\end{sloppypar}

 As follows from Eq. (\ref{18}), if $N=2$ then
\begin{equation}
d\rho^F(int)=\frac{d^2{\bf k}_{\bot}d\xi}{2(2\pi)^3\xi(1-\xi)}
=\frac{M_0({\bf k})d^3{\bf k}}{2(2\pi)^3\omega_1({\bf k})
\omega_2({\bf k})}
\label{23}
\end{equation}
where ${\bf k}\equiv {\bf k}_1=-{\bf k}_2$, $\xi=\xi_1=1-\xi_2$.
Let $int_i$ be a full set of the internal momentum variables
for the system $(1,...i-1,i+1,...N)$ and $d\rho^F(int_i)$ be
the internal volume element in the internal space $H_{int}^{(i)F}$ for
this system. Then, by analogy with the derivation of
Eq. (\ref{23}) from Eq. (\ref{18}), it is easy to show that
\begin{equation}
d\rho^F(int)=\frac{d^2{\bf k}_{i\bot}d\xi_i}{2(2\pi)^3\xi_i(1-\xi_i)}
d\rho^F(int_i)
\label{24}
\end{equation}
Therefore the normalization condition (\ref{19}) can be written as
\begin{eqnarray}
&&||\chi^F||_F^2= \sum_{\sigma_1...\sigma_N}\int\nolimits
|\chi^F({\bf k}_{i\bot},\xi_i,int_i,
\sigma_1,...\sigma_N)|^2 \cdot\nonumber\\
&&\frac{d^2{\bf k}_{i\bot}d\xi_i}{2(2\pi)^3\xi_i(1-\xi_i)}
d\rho^F(int_i)\, < \, \infty
\label{25}
\end{eqnarray}

If $M_i$ is the free mass operator for the system $(1,...i-1,i+1,...N)$
then, as follows from Eqs. (\ref{17}) and (\ref{22}),
\begin{eqnarray}
&&M_0\equiv M_0({\bf k}_i,M_i)\equiv M_0({\bf k}_{i\bot},\xi_i,M_i)=
\omega_i({\bf k}_i)+\nonumber\\
&&(M_i^2+{\bf k}_i^2)^{1/2}=[\frac{m_i^2}{\xi_i}+
\frac{M_i^2}{1-\xi_i}+\frac{{\bf k}_{i\bot}^2}{\xi_i(1-\xi_i)}]^{1/2}
\label{26}
\end{eqnarray}

\begin{sloppypar}
 Instead of $({\bf P}_{\bot},P^+)$ it is also possible to choose
$({\bf G}_{\bot},G^+)$ as the external variables while the internal
variables can be chosen as above. Now we introduce $H_{int}^P$ as the
space of functions
$\chi^P({\bf k}_1,\sigma_1,...{\bf k}_N,\sigma_N)$ such that
\begin{equation}
||\chi^P||_P^2=\sum_{\sigma_1...\sigma_N} \int\nolimits
|\chi^P({\bf k}_1,\sigma_1,...{\bf k}_N,\sigma_N)|^2
d\rho^P(int)\, <\, \infty
\label{27}
\end{equation}
where
\begin{equation}
d\rho^P(int)=M_0^2d\rho^F(int)
\label{28}
\end{equation}
Then by analogy with Eq. (\ref{20}) it is easy to show that the
Hilbert space of functions satisfying Eq. (\ref{14}) can be realized
as the space of functions $\varphi({\bf G}_{\bot},G^+)$ with the range
in $H_{int}^P$ and such that
\begin{equation}
\int\nolimits ||\varphi({\bf G}_{\bot},G^+)||_P^2 d\rho({\bf G}_{\bot},
G^+)\, <\, \infty
\label{29}
\end{equation}
\end{sloppypar}

\begin{sloppypar}
 A direct calculation shows that in the variables $({\bf G}_{\bot},
G^+,{\bf k}_1,...{\bf k}_N)$ the representation generators of the
Poincare group have the form (compare with Eq. (\ref{21}))
\begin{eqnarray}
&&P^+=M_0G^+,\quad {\bf P}_{\bot}=M_0{\bf G}_{\bot},\quad
P^-=M_0 G^-=M_0\frac{1+{\bf G}_{\bot}^2}{2G^+},\nonumber\\
&&M^{+-}=\imath G^+\frac{\partial}{\partial G^+},
\quad M^{+j}=-\imath G^+\frac{\partial}{\partial G^j},\nonumber\\
&& M^{xy}=l^z({\bf G}_{\bot})+S^z, \quad
M^{-j}=-\imath(G^j\frac{\partial}{\partial G^+}+
G^-\frac{\partial}{\partial G^j})- \nonumber\\
&&\frac{\epsilon_{jl}}{G^+}(S^l+G^lS^z)
\label{30}
\end{eqnarray}
\end{sloppypar}

\begin{sloppypar}
 If the particles interact with each other then one of the simplest way
to preserve the relativistic commutation relations (\ref{1}) is to
replace $M_0$ in Eq. (\ref{21}) by the mass operator ${\hat M}^F$
which acts only through the variables of the space $H_{int}^F$ and
commutes with ${\bf S}$. Then the representation generators of the
Poincare group obviously have the form
\begin{eqnarray}
&&P^+=P^+,\quad {\bf P}_{\bot}={\bf P}_{\bot},\quad {\hat P}^-=
\frac{({\hat M}^F)^2+{\bf P}_{\bot}^2}{2P^+}, \nonumber\\
&&M^{+-}=\imath P^+\frac{\partial}{\partial P^+},\quad
M^{+j}=-\imath P^+\frac{\partial}{\partial P^j},\nonumber\\
&&M^{xy}=l^z({\bf P}_{\bot})+S^z, \quad
{\hat M}^{-j}=-\imath(P^j\frac{\partial}{\partial P^+}+
{\hat P}^-\frac{\partial}{\partial P^j})-\nonumber\\
&&\frac{\epsilon_{jl}}{P^+}({\hat M}^FS^l+P^lS^z)
\label{31}
\end{eqnarray}
Such a procedure was first proposed by Bakamdjian and Thomas \cite{BT}.
It is obvious that the generators in Eq. (\ref{31}) are given in the
front form.
\end{sloppypar}

 Analogously, we can
replace $M_0$ in Eq. (\ref{30}) by the mass operator ${\hat M}^P$
which acts only through the variables of the space $H_{int}^P$ and
commutes with ${\bf S}$. Then the representation generators of the
Poincare group obviously have the form
\begin{eqnarray}
&&{\hat P}^+={\hat M}^PG^+,\quad {\hat{\bf P}}_{\bot}=
{\hat M}^P{\bf G}_{\bot},\quad {\hat P}^-={\hat M}^P G^-=\nonumber\\
&&{\hat M}^P\frac{1+{\bf G}_{\bot}^2}{2G^+},\quad
M^{+-}=\imath G^+\frac{\partial}{\partial G^+},
\quad M^{+j}=-\imath G^+\frac{\partial}{\partial G^j},\nonumber\\
&&M^{xy}=l^z({\bf G}_{\bot})+S^z,\quad
M^{-j}=-\imath(G^j\frac{\partial}{\partial G^+}+
G^-\frac{\partial}{\partial G^j})-\nonumber\\
&&\frac{\epsilon_{jl}}{G^+}(S^l+G^lS^z)
\label{32}
\end{eqnarray}
The generators in this expression are obviously given in the point form.

 If $\Gamma_i^F$ and $\Gamma_i^P$ $(i=1,...10)$ are the generators
given by Eqs. (\ref{31}) and (\ref{32}) respectively, then, as shown by
Sokolov and Shatny \cite{SoSh}, if ${\hat M}^F$ and ${\hat M}^P$ are
unitarily equivalent, the sets $\Gamma_i^F$ and $\Gamma_i^P$ are
unitarily equivalent too.

 In addition to extended relativistic invariance, the generators should
also satisfy cluster separability (see Refs.
\cite{sok,CP,M} for details). In the framework of the method of packing
operator developed by Sokolov \cite{sok} it can be shown
\cite{sok,CP,lev1} that the most general
form of the generators in the front and point forms is
$A^F\Gamma_i^F (A^F)^{-1}$ and $A^P\Gamma_i^P (A^P)^{-1}$ respectively
where $A^F$ and $A^P$ are some unitary operators.

 We shall consider a simple model when ${\hat M}^P=M_0+v_N^P$ where $v_N^P$
is the fully linked part of the mass operator. In other words, there
is only the $N$-body interaction and there are no $N'$-body interactions
if $N'<N$. Then relativistic invariance and cluster separability are
obviously satisfied if $[v_N^P,{\bf S}]=0$ and $A^P=1$. Analogously we
can consider the case when ${\hat M}^F=M_0+v_N^F$, $[v_N^F,{\bf S}]=0$
and $A^F=1$.

 Of course, the real mass operator contains all $N'$-body interactions
and $A^P\neq 1$. However, it is reasonable to assume that the nucleon
binding energy and wave function are mainly defined by some confining
interaction $v_N$ which cannot be determined in the framework of
perturbative QCD, while the other interaction operators in ${\hat M}^P$
can be determined in this framework. Such a situation takes place in
many realistic models which successfully describe the baryon spectra
(see e.g. Ref. \cite{CI}).

\section{Parton model as a consequence of IA
in the front form of dynamics}
\label{S4}

 For systems with a finite number of particles the operator
${\hat J}^{\mu}(x)$ as a usual operator valued function of $x$ is well
defined \cite{lev} and, as follows from Eqs. (\ref{1}), (\ref{3})
and (\ref{4}), if the operator ${\hat P}$ is known then
${\hat J}^{\mu}(x)$ is fully defined by the operator ${\hat J}^{\mu}(0)$
satisfying Eq. (\ref{6}).

 We shall always assume that all particles having the electric charge
are structureless and their spin is equal to 1/2. Then the one-particle
current operator
for particle $i$ acts over the variables of this particle as
\begin{eqnarray}
J_i^{\mu}(0)\varphi({\bf p}_{i\bot},p_i^+,\sigma_i)&=&
r_i\sum_{\sigma_i'}\int\nolimits [\bar{w}_i(p_i,\sigma_i)\gamma^{\mu}
w_i(p_i',\sigma_i')]\cdot\nonumber\\
&&\varphi({\bf p}_{i\bot}',p_i^{'+},
\sigma_i')d\rho({\bf p}_{i\bot}',p_i^{'+})
\label{33}
\end{eqnarray}
and over the variables of other particles it acts as the identity
operator. Here $r_i=e_i/e_0$ is the ratio of the particle electric
charge to the unit electric charge, $w_i(p_i,\sigma_i)$ is the Dirac light
cone spinor, $\gamma^{\mu}$ is the
Dirac $\gamma$-matrix, and $\bar{w}=w^{+}\gamma^0$. The form of
$w_i(p_i,\sigma_i)$ in the spinor representation of the Dirac
$\gamma$-matrices is
\begin{equation}
w_i(p_i,\sigma_i)=\sqrt{m_i}
\left\|\begin{array}{c}
\beta({\bf p}_{i\bot}/m_i,p_i^+/m_i)\chi(\sigma_i)\\
\beta({\bf p}_{i\bot}/m_i,p_i^+/m_i)^{-1+}\chi(\sigma_i)
\end{array}\right\|
\label{34}
\end{equation}
where $\chi(\sigma)$ is the ordinary spinor describing the state with
the spin projection on the $z$ axis equal to $\sigma$.

\begin{sloppypar}
 By definition, the operator ${\hat J}^{\mu}(0)$ in IA is given by
\begin{equation}
{\hat J}^{\mu}(0)=J^{\mu}(0)=\sum_{i=1}^{N}J_i^{\mu}(0)
\label{35}
\end{equation}
 In the front form such a current operator satisfies neither Lorentz
invariance nor invariance under the discrete symmetries. The
violation of Lorentz invariance has been explained in Sect. \ref{S2}
while the violation of the discrete symmetries follows from the
relations
\begin{eqnarray}
&&{\hat U}_P({\hat J}^0(0),{\hat {\bf J}}(0)){\hat U}_P^{-1}=
({\hat J}^0(0),-{\hat {\bf J}}(0))=\nonumber\\
&&{\hat U}_T({\hat J}^0(0),{\hat {\bf J}}(0)){\hat U}_T^{-1}
\label{36}
\end{eqnarray}
(recall that ${\hat U}_P$ and ${\hat U}_T$ in the front form are
necessarily interaction dependent). Nevertheless in this section we
carry out the calculations in IA for ${\hat J}^{\mu}(0)$ in the
front form and show that the results are exactly the same as in the
parton model. The analogous calculations have been carried out
elsewhere (see e.g. Ref. \cite{part,Web} and references therein).
\end{sloppypar}

 As follows from Eqs. (\ref{3}) and (\ref{10}), the hadronic tensor
can be written in the form
\begin{eqnarray}
W^{\mu\nu}&=&\frac{1}{4\pi}\sum_{X}(2\pi)^4\delta^{(4)}(P'+q-P)
\langle P',\chi^F|\nonumber\\
&&J^{\mu}(0)|X\rangle \langle X|J^{\nu}(0)|P',\chi^F\rangle
\label{37}
\end{eqnarray}
where the sum is taken over all possible final states $|X\rangle$,
and $P$ is the four-momentum of the state $|X\rangle$. We will
calculate this tensor only in the Bjorken limit when $Q^2$ and
$P'q$ are very large but the Bjorken variable $x=Q^2/2(P'q)$ is not
too close to 0 or 1 (from now on we will use $x$ only to denote the
Bjorken variable).

 We assume as usual (see e.g. Refs.
\cite{ER,EFP,Saw,Gurv}) that the final state interaction of the struck
quark with the remnants of the target is the effect of
order $(m/Q)^2$. Then in the model for the mass operator considered in
Sect. \ref{S3}, we can write $|X\rangle$ as the states of $N$ free
particles with the 4-momenta $p_i"$ and spin projections $\sigma_i"$
$(i=1,...N)$:
\begin{equation}
|X\rangle=\prod_{i=1}^{N}|p_i",\sigma_i"\rangle
\label{38}
\end{equation}
Taking into account the normalization of free states in the scattering
theory we write the wave functions of these states in the form
\begin{equation}
|p_i",\sigma_i"\rangle =2(2\pi)^3p_i^{"+}\delta^{(2)}({\bf p}_{i\bot}-
{\bf p}_{i\bot}")\delta(p_i^+-p_i^{"+})\delta_{\sigma_i\sigma_i"}
\label{39}
\end{equation}
where $\delta_{\sigma_i\sigma_i"}$ is the Cronecker symbol.

 Analogously if the Poincare group generators are given by
Eq. (\ref{31}), the wave function of the initial nucleon can be written as
\begin{equation}
|P',\chi^F\rangle = 2(2\pi)^3P^{'+}
\delta^{(2)}({\bf P}_{\bot}-{\bf P}_{\bot}')\delta(P^+-P^{'+})\chi^F
\label{40}
\end{equation}
where the internal nucleon wave function $\chi^F$ is normalized as
$||\chi^F||_F=1$.

\begin{sloppypar}
 From now on it will be convenient to denote the momenta of final
particles without two primes, i.e. as $p_1,...p_N$. Then, as follows
from Eqs. (\ref{33}), (\ref{35}) and (\ref{37}-\ref{40}),
\begin{eqnarray}
&&W^{\mu\nu}=\frac{1}{4\pi}\sum_{\sigma_1...\sigma_N}\sum_{i,j=1}^{N}
\sum_{\sigma_i',\sigma_j'}\int\nolimits \{\prod_{l=1}^{N}
d\rho({\bf p}_{l\bot},p_l^+)\}\frac{r_ir_j}{\xi_i'\xi_j'}\cdot\nonumber\\
&&(2\pi)^4\delta^{(4)}(P'+q-P)\langle \chi^F({\bf k}_j',int_j,
\sigma_1,...\sigma_j',...\sigma_N)|\nonumber\\
&&[\bar{w}_j(p_j',\sigma_j')\gamma^{\mu}w_j(p_j,\sigma_j)]
[\bar{w}_i(p_i,\sigma_i)\gamma^{\nu}
w_i(p_i',\sigma_i')]\nonumber\\
&&|\chi^F({\bf k}_i',int_i,\sigma_1,...\sigma_i',...\sigma_N)\rangle
\label{41}
\end{eqnarray}
where $p_i'$, ${\bf k}_i'$ and $\xi_i'$ are defined as follows.
\end{sloppypar}

 If $P_i=p_1+...+p_{i-1}+p_{i+1}+...p_N$ then the four-vector $p_i'$ is
such that $p_i^{'2}=m_i^2$, $p_i^{'+}=P^{'+}-P_i^+$,
${\bf p}_{i\bot}'={\bf P}_{\bot}'-{\bf P}_{i\bot}$. Then (compare with
Eqs. (\ref{17}) and (\ref{22}))
\begin{eqnarray}
k_i'&=&L[\beta(\frac{{\bf P}_{\bot}'}{M_0({\bf k}_i',M_i)},
\frac{P^{'+}}{M_0({\bf k}_i',M_i)})]^{-1}p_i',\nonumber\\
&&\xi_i'=\frac{\omega_i({\bf k}_i')+k_i^{'z}}{M_0({\bf k}_i',M_i)}
\label{42}
\end{eqnarray}
and $k_i'=(\omega_i({\bf k}_i'),{\bf k}_i')$.

 We can write $P_i$ on the one hand as a function of $P',p_i',int_i$
and on the other hand as a function of $P,p_i,int_i$. Therefore
(compare with Eq. (\ref{17})) ${\bf k}_i'$ can be defined by the condition
\begin{eqnarray}
&&L[\beta(\frac{{\bf P}_{\bot}'}{M_0({\bf k}_i',M_i)},
\frac{P^{'+}}{M_0({\bf k}_i',M_i)})]((M_i^2+{\bf k}_i^{'2})^{1/2},
-{\bf k}_i')=\nonumber\\
&&L[\beta(\frac{{\bf P}_{\bot}}{M_0({\bf k}_i,M_i)},
\frac{P^+}{M_0({\bf k}_i,M_i)})]((M_i^2+{\bf k}_i^2)^{1/2},-{\bf k}_i)
\label{43}
\end{eqnarray}

 As follows from Eq. (\ref{41}), the mass $M=M_0({\bf k}_i,M_i)$ of the
final state satisfies
the condition $M^2=(P'+q)^2$. Hence
\begin{equation}
M^2=m^2+\frac{Q^2(1-x)}{x}
\label{44}
\end{equation}
and in the Bjorken limit $m^2$ in this expression can be neglected.

 It is convenient to consider the process in the IMF where
${\bf P}_{\bot}'={\bf q}_{\bot}=0$, and $P^{'z}$ is positive and very
large. By analogy with the Breit frame for elastic processes we choose
the reference frame in which ${\bf P}+{\bf P}'=0$. It is easy to show
that in this reference frame
\begin{eqnarray}
&&q^0=2|{\bf P}'|(1-x),\quad P^{'+}=\sqrt{2}|{\bf P}'|,\quad
q^+=-\sqrt{2}|{\bf P}'|x, \nonumber\\
&&P^+=\sqrt{2}|{\bf P}'|(1-x)
\label{45}
\end{eqnarray}
Then as follows from Eqs. (\ref{16}), (\ref{22}), (\ref{43}) and
(\ref{45})
\begin{equation}
{\bf k}_{i\bot}'={\bf k}_{i\bot},\quad
\xi_i'=(1-x)\xi_i+x
\label{46}
\end{equation}

 We assume that the internal wave function $\chi^F$ effectively
cuts the contribution of large momenta, and therefore the contribution
to the integrals containing $\chi^F({\bf k}_i',int_i)$ is given only by the
momenta with $|{\bf k}_i'|\leq m_0$, $|{\bf k}_l|\leq m_0$ $(l=1,...i-1,
i+1,...N)$ where $m_0$ is some parameter satisfying the condition
$m_0 \ll Q$. In turn, as follows from Eq. (\ref{42}), this implies that
$\xi_i'$ should not be close to 0 or 1, and, as follows from
Eq. (\ref{26}), in the Bjorken limit $|k_i^z|\approx M/2$. If $k^z>0$ then,
as follows from Eqs. (\ref{22}) and (\ref{26}), $\xi_i\approx 1$.
However in this case $\xi_i'$ is close to 1 as follows from Eq. (\ref{46}).
Therefore the only possibility is $k^z<0$, $\xi_i\approx 0$. Then, as
follows from Eq. (\ref{46}), in the Bjorken limit $\xi_i'=x$.
Analogously in the Bjorken limit the contribution to the integrals
containing $\chi^F({\bf k}_j',int_j)$ is not negligible only if
$|{\bf k}_j'|\leq m_0$, $|{\bf k}_n|\leq m_0$ $(n=1,...j-1,
j+1,...N)$, $|k_j^z|\approx M/2$ and $k_j^z<0$. All these conditions
are compatible with each other only if $i=j$, i.e. the quarks absorb the
virtual photon incoherently.  As follows from Eqs. (\ref{22}) and
(\ref{42}), the result $\xi_i'=x$ fully agrees with the interpretation
of the quantity $x$ in the parton model as the momentum fraction of the
struck quark in the IMF.

 As follows from Eqs. (\ref{20}), (\ref{24}) and (\ref{26}), if
the above conditions are satisfied then
\begin{equation}
(2\pi)^4\delta^{(4)}(P'+q-P)\prod_{i=1}^{N}
d\rho({\bf p}_{i\bot},p_i^+)=\frac{d^2{\bf k}_{i\bot}}{8\pi^2M^2}
d\rho^F(int_i)
\label{47}
\end{equation}
Therefore, as follows from Eqs. (\ref{17}), (\ref{34}), (\ref{41}),
and (\ref{45}-\ref{47}), the final result for the hadronic tensor is
\begin{eqnarray}
W^{\mu\nu}&=&\sum_{i=1}^{N} r_i^2 \int\nolimits \langle
\chi^F({\bf k}_{i\bot},\xi_i'=x,int_i)|S_i^{\mu\nu}|\nonumber\\
&&\chi^F({\bf k}_{i\bot},\xi_i'=x,int_i)\rangle
\frac{d^2{\bf k}_{i\bot}d\rho^F(int_i)}{4(2\pi)^3x(1-x)}
\label{48}
\end{eqnarray}
where we do not write the spin variables in the arguments of the
function $\chi^F$, the scalar product is taken over these variables, and
the tensor operator $S_i^{\mu\nu}$ is as follows. It is equal to zero
if either $\mu$ or $\nu$ is equal to $\pm$, while if $j,l=x,y$ then
$S_i^{jl}=\delta_{jl}+2\imath \epsilon_{jl} s_i^z$, where
$s_i^z$ is the $z$ component of the spin operator for particle $i$.

 Let us introduce the notation
\begin{equation}
\rho_i(x)= \sum_{\sigma_1...\sigma_N} \int\nolimits
|\chi^F({\bf k}_{i\bot},\xi_i'=x,int_i,\sigma_1,...\sigma_N)|^2
\frac{d^2{\bf k}_{i\bot}d\rho^F(int_i)}{2(2\pi)^3x(1-x)}
\label{49}
\end{equation}
This quantity is called the parton density since, as follows from
Eq. (\ref{25}), $\rho_i(\xi_i')d\xi_i'$ is the probability of the event
that particle $i$ in the bound state has the value of $\xi_i'$ in the
interval $(\xi_i',\xi_i'+d\xi_i')$.

 It is well-known that the average value of the hadronic tensor over
all initial spin states is equal to
\begin{eqnarray}
&&W^{\mu\nu}(P',q)=(\frac{q^{\mu}q^{\nu}}{q^2}-\eta^{\mu\nu})
F_1(x,q^2)+\nonumber\\
&&\frac{1}{(P'q)}(P^{'\mu}-\frac{q^{\mu}(P'q)}{q^2})
(P^{'\nu}-\frac{q^{\nu}(P'q)}{q^2})F_2(x,q^2),
\label{50}
\end{eqnarray}
Then, as follows from Eqs. (\ref{48}) and (\ref{49}), the structure
functions $F_1$ and $F_2$ depend only on $x$
(this phenomenon is known as Bjorken Scaling):
\begin{equation}
F_1(x)=\frac{1}{2}\sum_{i=1}^{N}r_i^2\rho_i(x),\quad F_2(x)=2xF_1(x)
\label{51}
\end{equation}
(the last equality is known as the Callan-Gross relation \cite{CallGr}).
These expressions for the structure functions have been derived by many
authors in the framework of the parton model. Equation (\ref{48}) also
makes it possible to write the expression for the polarized structure
functions, but we shall not dwell on this question.

 One might think that the above results are natural since they fully
agree with the parton model. However the following question arises. Since
the current in IA does not satisfy Lorentz invariance, P invariance
and T invariance, the results for the structure functions depend on the
reference frame in which these functions are calculated. An argument in
favor of choosing the IMF is that in this reference frame the
current conservation in the Bjorken limit is restored since the tensor
$W^{\mu\nu}$ given by Eq. (\ref{48}) satisfies the continuity equation
$q_{\mu}W^{\mu\nu}=q_{\nu}W^{\mu\nu}=0$.
Another well-known arguments are based on the approach proposed by
Weinberg \cite{Wein} and developed by several authors (see, for example,
Refs. \cite{BrLep,Namysl}). Let us note however that though quantum field
theory in the IMF seems natural and has some advantages, it also has some
serious difficulties which are not present in the usual formulation
\cite{GlWil}.

 In our opinion, a rather strange feature of the above results is as
follows. By looking through the derivation of these results one can
easily see that the initial state is treated in fact not as the bound
state but as the free state of noninteracting particles. Indeed, we have
never used the fact that the initial state is the eigenstate of the mass
operator ${\hat M}^F$ with the eigenvalue $m$: ${\hat M}^F\chi^F=m\chi^F$.
In IA in the front form the relation between the quantities
${\bf k}_i'$ and ${\bf k}_i$ (see Eq. (\ref{43})) is derived from the
condition that the four-vectors $(M_i^2+{\bf k}_i^{'2})^{1/2},
-{\bf k}_i')$
and $(M_i^2+{\bf k}_i^2)^{1/2},-{\bf k}_i)$ are connected by the
Lorentz boosts in the initial and final states. The problem whether
particle $i$ does not interact with the other particles in the final
state deserves a separate investigation and we will not discuss this
problem in the present paper, but it is strange that we neglect the
interaction in the initial
state and write the free mass $M_0({\bf k}_i',M_i)$ instead of the real
mass $m$ which has the initial state.

 The effect of binding can be explicitly taken into account in models
where the current operator satisfies relativistic invariance and current
conservation. This problem is considered in the next section.

 The unpolarized hadronic tensor for the neutrino (antineutrino) -
nucleon scattering has the form
\begin{eqnarray}
W^{\mu\nu}(P',q)&=&-\eta^{\mu\nu}F_1(x,q^2)+\frac{P^{'\mu}P^{'\nu}}{2(P'q)}
F_2(x,q^2)-\nonumber\\
&&\imath \epsilon^{\mu\nu\rho\lambda}\frac{P'_{\rho}q_{\lambda}}
{2(P'q)}F_3(x,q^2)
\label{52}
\end{eqnarray}
where $\epsilon^{\mu\nu\rho\lambda}$ is the fully antisymmetric tensor
with $\epsilon^{0123}=1$ and the terms with $q^{\mu}$ and $q^{\nu}$ are
dropped since they give zero after multiplication by the leptonic tensor.
The calculation of the deep inelastic scattering caused by the charged
weak currents can be carried out by analogy with the above calculation
and the result is
\begin{eqnarray}
&&F_1^{\nu p}(x)=\sum_{i=d,s,\bar{u},\bar{c}}\rho_i(x),\quad
F_1^{\bar{\nu} p}(x)=\sum_{i=u,c,\bar{d},\bar{s}}\rho_i(x),\nonumber\\
&&F_3^{\nu p}(x)=2\sum_{i=d,s}\rho_i(x)-
2\sum_{i=\bar{u},\bar{c}}\rho_i(x),\nonumber\\
&&F_3^{\bar{\nu} p}(x)=2\sum_{i=u,c}\rho_i(x)-
2\sum_{i=\bar{d},\bar{s}}\rho_i(x),\nonumber\\
&&F_2^{\nu p}(x)=2xF_1^{\nu p}(x),\quad
F_2^{\bar{\nu} p}(x)=2xF_1^{\bar{\nu} p}(x)
\label{53}
\end{eqnarray}
where we assume for simplicity that the proton does not contain $b$ and
$t$ quarks.

\begin{sloppypar}
\section{Consistent calculation of the hadronic tensor}
\label{S5}
\end{sloppypar}

 In this section we calculate the hadronic tensor assuming that
${\hat J}^{\mu}(0)=J^{\mu}(0)$ in the point form. Then extended
Poincare invariance is satisfied automatically (see Sects. \ref{S1}
and \ref{S2}) and if ${\hat J}^{\mu}(0)$ is free in some reference
frame it will remain free in all reference frames obtained from this
one by means of Lorentz boosts. As shown in Ref. \cite{lev}, the operator
${\hat J}^{\mu}(0)$ is fully defined by its matrix elements in
the reference frame where ${\bf G}'+{\bf G}=0$. Here $G'=P'/m$ is the
four-velocity of the system in the initial state and $G=P/M$ is the same
quantity in the final state.

 Since now the current operator satisfies Poincare invariance the
calculations can be carried out in any reference frame. Therefore
we can again suppose that ${\bf P}_{\bot}'={\bf q}_{\bot}=0$ and
$P^{'z}>0$. Therefore ${\bf G}_{\bot}'={\bf G}_{\bot}=0$, and $G^{'z}>0$.
If ${\bf G}+{\bf G}'=0$ then $G^z<0$. As follows from these
expressions and Eq. (\ref{44}), in the Bjorken limit
\begin{equation}
(G^z)^2=\frac{Q}{4m[x(1-x)]^{1/2}}
\label{54}
\end{equation}
Therefore $G^{'z}\gg 1$ in the Bjorken limit.

 As follows from Eq. (\ref{5}), in the reference frame under
consideration the $\bot$ components of ${\hat J}^{\mu}(0)$ are not
constrained by the continuity equation
while the longitudinal components should satisfy the condition
\begin{eqnarray}
&&(M-m)G^{'0}\langle X|{\hat J}^0(0)|P',\chi^P\rangle =\nonumber\\
&&-(M+m)G^{'z}\langle X|{\hat J}^z(0)|P',\chi^P\rangle
\label{55}
\end{eqnarray}
Therefore in the Bjorken limit
\begin{equation}
\langle X|{\hat J}^0(0)|P',\chi^P\rangle =
-\langle X|{\hat J}^z(0)|P',\chi^P\rangle
\label{56}
\end{equation}
Note that the operator $G$ is free by construction (see Eq. (\ref{32}))
and therefore the free operator $J^{\mu}(0)$ satisfies the condition
\begin{eqnarray}
&&G^{'0}\langle X|J^0(0)(M-M_0)|P',\chi^P\rangle =  \nonumber\\
&&-G^{'z}\langle X|J^z(0)(M+M_0)|P',\chi^P\rangle
\label{57}
\end{eqnarray}
Since $M\gg M_0$ in the Bjorken limit we conclude that the choice
${\hat J}^{\mu}(0)=J^{\mu}(0)$ is compatible with the continuity equation.

\begin{sloppypar}
 This choice is also compatible with macrolocality. Indeed, as
follows from Eqs. (\ref{3}) and (\ref{32})
\begin{eqnarray}
&&{\hat J}^{\mu}({\bf x}){\hat J}^{\nu}(0)\varphi (G)=
exp(-\imath {\hat M}^P({\bf G}{\bf x}))J^{\mu}(0)\cdot\nonumber\\
&& exp(\imath {\hat M}^P({\bf G}{\bf x}))J^{\nu}(0)\varphi(G)
\label{58}
\end{eqnarray}
Since the operator $J^{\mu}(0)$ is not singular, the strong limit of
${\hat J}^{\mu}({\bf x}){\hat J}^{\nu}(0)$ is equal to zero if
$|{\bf x}|\rightarrow \infty$. This can be proved, for example, by
analogy with the proof of space separability in Ref. \cite{sok1}.
Analogously it is easy to see that the
same is valid for the strong limit of
${\hat J}^{\nu}(0){\hat J}^{\mu}({\bf x})$.
\end{sloppypar}

 We conclude that the choice ${\hat J}^{\mu}(0)=J^{\mu}(0)$ in the
point form is consistent in the sense that it is compatible with
extended Poincare invariance, continuity equation and macrolocality.

 As follows from Eq. (\ref{32}) and the normalization of states in the
scattering theory, the wave function of the initial nucleon in the point
form can be written as
\begin{equation}
|P',\chi^P\rangle =\frac{2}{m}(2\pi)^3G^{'+}
\delta^{(2)}({\bf G}_{\bot}-{\bf G}_{\bot}')\delta(G^+-G^{'+})\chi^P
\label{59}
\end{equation}
where $||\chi^P||_P=1$.

\begin{sloppypar}
 Now the hadronic tensor has the form (compare with Eq. (\ref{37}))
\begin{eqnarray}
W^{\mu\nu}&=&\frac{1}{4\pi}\sum_{X}(2\pi)^4\delta^{(4)}(P'+q-P)
\langle P',\chi^P|\nonumber\\
&&J^{\mu}(0)|X\rangle \langle X|J^{\nu}(0)|P',\chi^P\rangle
\label{60}
\end{eqnarray}
We will calculate this tensor in the reference frame considered above.
As follows from Eq. (\ref{56}), in this reference frame the matrix
elements of the operator ${\hat J}^+(0)$ is negligible in comparison
with the matrix elements of the operator ${\hat J}^-(0)$. Therefore it is
sufficient to calculate the tensor $W^{\mu\nu}$ for $\mu,\nu=x,y,-$.
It is also easy to show that in this reference frame
\begin{equation}
{\bf q}_{\bot}=0,\quad
q^+=-\frac{x^{3/4}(mQ)^{1/2}}{\sqrt{2}(1-x)^{1/4}},
\quad q^-=\frac{(1-x)^{1/4}Q^{3/2}}{\sqrt{2}m^{1/2}x^{3/4}}
\label{61}
\end{equation}
Note that $q^-\gg |q^+|$.
\end{sloppypar}

 Let us define the four-vector $p_i'$ by the condition
\begin{equation}
p_i^{'2}=m_i^2,\quad \frac{p_i'+P_i}{|p_i'+P_i|}=G'
\label{62}
\end{equation}
Then a direct calculation shows that if $P_i$ is fixed then
\cite{sok}
\begin{equation}
d\rho({\bf p}_{i\bot}',p_i^{'+})=\frac{|p_i'+P_i|^4}{(p_i',p_i'+P_i)}
d\rho({\bf G}_{\bot}',G^{'+})
\label{63}
\end{equation}
Therefore, as follows from Eqs. (\ref{33}), (\ref{35}) and (\ref{38}),
(\ref{39}), (\ref{59}), (\ref{60}) and (\ref{62}),
\begin{eqnarray}
&&W^{\mu\nu}=\frac{1}{4\pi}\sum_{\sigma_1...\sigma_N}\sum_{i,j=1}^{N}
\sum_{\sigma_i',\sigma_j'}\int\nolimits \{\prod_{l=1}^{N}
d\rho({\bf p}_{l\bot},p_l^+)\}\frac{r_ir_j}{m^2}\cdot\nonumber\\
&&(2\pi)^4\delta^{(4)}(P'+q-P)\frac{M_0({\bf k}_i',M_i)^3
M_0({\bf k}_j',M_j)^3}{\omega_i({\bf k}_i')
\omega_j({\bf k}_j')}\cdot\nonumber\\
&&\langle \chi^P({\bf k}_j',int_j,\sigma_1,...\sigma_j',...\sigma_N)|
[\bar{w}_j(p_j',\sigma_j')\gamma^{\mu}w_j(p_j,\sigma_j)]\cdot\nonumber\\
&&[\bar{w}_i(p_i,\sigma_i)\gamma^{\nu}w_i(p_i',\sigma_i')]|
\chi^P({\bf k}_i',int_i,\sigma_1,...\sigma_i',...\sigma_N)\rangle
\label{64}
\end{eqnarray}
where (compare with Eq. (\ref{42}))
\begin{equation}
k_i'=L[\beta({\bf G}_{\bot}',G^{'+})]^{-1}p_i',\quad
\xi_i'=\frac{\omega_i({\bf k}_i')+k_i^{'z}}{M_0({\bf k}_i',M_i)}
\label{65}
\end{equation}
and $k_i'=(\omega_i({\bf k}_i'),{\bf k}_i')$.

  We can write $P_i$ on the one hand as a function of $G',p_i',int_i$
and on the other hand as a function of $G,p_i,int_i$. Therefore
(compare with Eq. (\ref{43})) ${\bf k}_i'$ can be defined by the condition
\begin{eqnarray}
&&L[\beta({\bf G}_{\bot}',G^{'+})]((M_i^2+{\bf k}_i^{'2})^{1/2},
-{\bf k}_i')=\nonumber\\
&&L[\beta({\bf G}_{\bot},G^+)]((M_i^2+{\bf k}_i^2)^{1/2},-{\bf k}_i)
\label{66}
\end{eqnarray}
As follows from Eq. (\ref{34}), in the reference frame under
consideration
\begin{equation}
{\bf k}_{i\bot}'={\bf k}_{i\bot},\quad k_i^{'z}=(1+2|G^z|^2)k_i^z-
2G^0G^z(M_i^2+{\bf k}_i^2)^{1/2}
\label{67}
\end{equation}
As follows from this expression, in the Bjorken limit
\begin{equation}
k_i^{'z}=(1+2|G^z|^2)[k_i^z+(M_i^2+{\bf k}_i^2)^{1/2}]-
\frac{(M_i^2+{\bf k}_i^2)^{1/2}}{4|G^z|^2}
\label{68}
\end{equation}
Therefore taking into account Eq. (\ref{26}) we conclude that
$|k_i^{'z}|\leq m_0$ (see Sect. \ref{S4}) if $|k_i^z|\approx M/2$
and $k_i^z<0$. Then, as follows from Eqs. (\ref{54}), (\ref{67})
and (\ref{68})
\begin{equation}
k_i^{'z}=\frac{M_i^2+{\bf k}_{i\bot}^2}{2m(1-x)}-\frac{m(1-x)}{2}
\label{69}
\end{equation}

 As follows from Eqs. (\ref{66}) and (\ref{69}),
\begin{equation}
M_0({\bf k}_{i\bot},\xi_i',M_i)(1-\xi_i')=m(1-x)
\label{70}
\end{equation}
where the function $M_0({\bf k}_{i\bot},\xi_i',M_i)$ is defined by
Eq. (\ref{26}).

\begin{sloppypar}
As follows from Eqs. (\ref{26}) and (\ref{70}), the explicit
expression of $\xi_i'$ as a function of ${\bf k}_{i\bot},M_i,x$ is
\begin{eqnarray}
\xi_i'=
\left\{\begin{array}{c}
\frac{1}{2}\{1-\alpha_i-\beta_i+[(1-\alpha_i-\beta_i)^2+
4\alpha_i]^{1/2}\} \\
\frac{1}{2}\{1-\alpha_i-\beta_i-[(1-\alpha_i-\beta_i)^2+
4\alpha_i]^{1/2}\}\\
\frac{m_i^2+{\bf k}_{i\bot}^2}{m_i^2+
{\bf k}_{i\bot}^2+M^{'2}(1-x)^2}
\end{array}\right.
\label{71}
\end{eqnarray}
if $M_i>m_i$, $M_i<m_i$ and $M_i=m_i$ respectively where
\begin{equation}
\alpha_i=\frac{m_i^2+{\bf k}_{i\bot}^2}{M_i^2-m_i^2},\quad
\beta_i=\frac{M^{'2}(1-x)^2}{M_i^2-m_i^2}
\label{72}
\end{equation}
Since $x\in [0,1]$, it follows from Eqs. (\ref{71}) and (\ref{72}) that
$\xi_i' \in [\xi_i^{min},1]$ where $\xi_i^{min}=\xi_i^{min}({\bf
k}_{i\bot},M_i)$ is a function of ${\bf k}_{i\bot},M_i$ which can be
determined from Eqs. (\ref{71}) and (\ref{72}) at $x=0$. It is easy to
see that $0<\xi_i^{min}<1$.
\end{sloppypar}

 By analogy with the consideration in the preceding section we can
show that in our model the quarks absorb the virtual photon incoherently.
Therefore
everything is ready for the calculation of the hadronic tensor in
Eq. (\ref{64}). Since we wish to compare the results with those obtained
in the front form, we note that, as follows from Eqs.
(\ref{25}), (\ref{27}) and (\ref{28}),
$\chi^P=\chi^F/M_0$. Then a simple calculation using Eqs. (\ref{17}),
(\ref{24}), (\ref{28}), (\ref{34}), (\ref{47}), (\ref{54}), (\ref{65}),
(\ref{67}) and (\ref{70})
shows that in the reference frame under consideration
\begin{eqnarray}
W^{\mu\nu}&=&\sum_{i=1}^{N} r_i^2 \int\nolimits \langle
\chi^F({\bf k}_{i\bot},\xi_i',int_i)|S_i^{\mu\nu}|
\chi^F({\bf k}_{i\bot},\xi_i',int_i)\rangle \cdot\nonumber\\
&&(\frac{1-x}{1-\xi_i'})^3[1+\frac{k_i^{'z}}{\omega_i({\bf k}_i')}]^2
\frac{d^2{\bf k}_{i\bot}d\rho^F(int_i)}{4(2\pi)^3\xi_i'(1-x)}
\label{73}
\end{eqnarray}
where $\xi_i'=\xi_i'({\bf k}_{i\bot},M_i,x)$ is given by Eqs.
(\ref{71}) and (\ref{72}).

 Let us introduce the function
\begin{eqnarray}
\tilde{\rho}_i(x)&=&\sum_{\sigma_1...\sigma_N} \int\nolimits
|\chi^F({\bf k}_{i\bot},\xi_i',
int_i,\sigma_1,...\sigma_N)|^2 \cdot\nonumber\\
&&(\frac{1-x}{1-\xi_i'})^3[1+\frac{k_i^{'z}}{\omega_i({\bf k}_i')}]^2
\frac{d^2{\bf k}_{i\bot}
d\rho^F(int_i)}{2(2\pi)^3\xi_i(1-x)}
\label{74}
\end{eqnarray}
In contrast with the function $\rho_i(x)$ defined by Eq. (\ref{49}),
the function $\tilde{\rho}_i(x)$ obviously has no probabilistic
interpretation. As follows from Eqs. (\ref{50}), (\ref{61}) and
(\ref{73}), in our model Bjorken Scaling and the Callan-Gross relation
\cite{CallGr} also take place since
\begin{equation}
F_1(x)=\frac{1}{2}\sum_{i=1}^{N} r_i^2 \tilde{\rho}_i(x),
\quad F_2(x)=2xF_1(x)
\label{75}
\end{equation}

 Analogously we can calculate the hadronic tensor for the neutrino
(antineutrino) - nucleon scattering, and, instead of Eq. (\ref{53}),
the result is
\begin{eqnarray}
&&F_1^{\nu p}(x)=\sum_{i=d,s,\bar{u},\bar{c}}\tilde{\rho}_i(x),\quad
F_1^{\bar{\nu} p}(x)=\sum_{i=u,c,\bar{d},
\bar{s}}\tilde{\rho}_i(x),\nonumber\\
&&F_3^{\nu p}(x)=2\sum_{i=d,s}\tilde{\rho}_i(x)-
2\sum_{i=\bar{u},\bar{c}}\tilde{\rho}_i(x),\nonumber\\
&&F_3^{\bar{\nu} p}(x)=2\sum_{i=u,c}\tilde{\rho}_i(x)-
2\sum_{i=\bar{d},\bar{s}}\tilde{\rho}_i(x),\nonumber\\
&&F_2^{\nu p}(x)=2xF_1^{\nu p}(x),\quad
F_2^{\bar{\nu} p}(x)=2xF_1^{\bar{\nu} p}(x)
\label{76}
\end{eqnarray}

 In connection with the discussion of the role of off-shellness in
Sect. \ref{S2}, let us note that
in our approach the four-momenta of interacting particles are
always on-mass shell (by analogy with the "old-fashioned" time
ordered perturbation theory) while  in the Feynman diagram approach
such four-momenta ${\tilde p}_i$ can be off-shell, i.e. in the
general case ${\tilde p}_i^2\neq m_i^2$. In diagrams which in the
conventional approach are supposed to be dominant ("handbag diagrams")
the four-momentum of the struck quark in the initial state is often
written in the form ${\tilde p}_i=x_iP'+{\tilde k}_i$ where all the
components of the vector ${\tilde k}_i$ are much smaller than
$Q$. Therefore the four-momentum of this quark in the final state
is equal to ${\tilde p}_i+q$ and the condition that the off-shellness
is much smaller than $Q^2$ implies that $|{\tilde p}_i^2|,
|({\tilde p}_i+q)^2|\ll Q^2$. Hence we easily derive that $x_i=x$ in
the Bjorken limit. It is obvious that $x_i=\tilde{p_i}^+/P^{'+}$ in
the Bjorken limit and this is the argument that the Bjorken variable
$x$ can be interpreted as the momentum fraction in the IMF.

 It is clear from Eq. (\ref{70}) that the equality $\xi_i'=x$ takes place
only if one neglects the difference between the free mass and the mass
of the bound state while in the general case $\xi_i'\neq x$. In the point
form $\xi_i'$ is not equal to the quantity $x_i$ since (see
Eq. (\ref{22})) the quantity $P^{'+}$ in the point form is not the same
as for free particles. Therefore the fact that $\xi_i'\neq x$ does not
contradict the relation $|{\tilde p}_i^2|\ll Q^2$. Let us
calculate the quantity ${\tilde p}_i^2-m_i^2$ in the parton model and
in our one. In both cases ${\tilde p}_i=P'-P_i$ and we take into
account that the vector $P_i$ is on-shell, i.e. $P_i^2=M_i^2$. As
follows from Eqs. (\ref{46}) and (\ref{67}), in both cases the $\bot$
components of the vector $P_i$ is equal to $-{\bf k}_{i\bot}$.
Therefore in both cases $P_i^-=(M_i^2+{\bf k}_{i\bot}^2)/2P_i^+$. In the
parton model $P_i$ is given by the left-hand-side of Eq. (\ref{43}),
and, as follows from Eqs. (\ref{42}), (\ref{46}) and the condition
$\xi_i'=x$, $P_i^+=(1-x)P^{'+}$. In our model
$P_i$ is given by the left-hand-side of Eq. (\ref{66}),
and, as follows from the conditions $G'=P'/m$ and Eqs. (\ref{65}),
(\ref{67}) and (\ref{70}), again $P_i^+=(1-x)P^{'+}$. Therefore a
simple calculation using Eq. (\ref{26}) shows that in both cases
\begin{equation}
\tilde{p}_i^2-m_i^2=x[m^2-M_0({\bf k}_{i\bot},x,M_i)^2]
\label{77}
\end{equation}

 On the other hand, let us calculate the quantity $q_i=p_i-p_i'$
which has the meaning of the four-momentum transferred to the
struck quark. Since $P'+q=P$ and $P=P_i+p_i$, we have
$q_i=q+P'-P_i-p_i'$.
Since in the front form the $+,\bot$ components of $P'$ are the same as
for noninteracting particles it is obvious that $q_i^+=q^+$ and
${\bf q}_{i\bot}={\bf q}_{\bot}$. At the same time, in the reference
frames considered in the preceding and this sections the minus
components of $P'$, $P_i$ and $p_i'$ are infinitely small. Therefore,
in the parton model $q_i=q$ and in our one $q_i^-=q^-$. Therefore,
as follows from Eqs. (\ref{17}), (\ref{54}), (\ref{61}), (\ref{67})
and (\ref{70}), in our model
\begin{equation}
{\bf q}_{i\bot}=0,\quad q_i^+=q^+\frac{(1-x)\xi_i'}{(1-\xi_i')x},\quad
q_i^-=q^-,\quad q_i^2=q^2 \frac{(1-x)\xi_i'}{(1-\xi_i')x}
\label{78}
\end{equation}

In the conventional theory we must have $q_i=q$
since the inequality of these quantities is prohibited by the
factorization theorem \cite{ER,EFP,CSS,Sterman} which (see the
discussion in
Sect. \ref{S2}) asserts in particular that upper parts of Feynman
diagrams dominating in DIS are fully defined by the hard part of the DIS
process. Roughly speaking this implies that the vertex describing
absorption of the virtual photon by the struck constituent
is not surrounded by virtual particles and therefore it is obvious
that $q_i=q$. However, as argued in Sect. \ref{S2}, the factorization
theorem does not take place in the general case.
Since in relativistic models the values of $M_0$ essentially
differ from $m$, it is obvious from Eq. (\ref{77}) that the
off-shellness is important. For this reason it is not clear why the
vertex describing the absorption of the virtual photon can be
considered in perturbation theory.

 Let us now consider a question which seems to be important for
understanding the difference between our model and the parton
one. In both cases the nucleon is described by the wave function
$\varphi(p_1,...p_N)$ where the spin variables are dropped. The
quantity $p_i$ is the four-momentum of {\it free} particle $i$. The
question arises whether $p_i$ can be interpreted as the
four-momentum of particle $i$ inside the nucleon when the particles
interact with each other. As pointed out by Coester \cite{coes2},
such an interpretation implies that
\begin{equation}
[{\hat P},p_i]=0,\quad {\hat U}(l)^{-1}p_i^{\mu}{\hat U}(l)=
L(l)^{\mu}_{\nu}p_i^{\nu}
\label{79}
\end{equation}
where ${\hat U}(l)$ is the representation operator corresponding to
$l\in SL(2,C)$. It is sufficient to satisfy these relations not on
the whole Hilbert space
but only on the subspace $H_0$ consisting of all possible single-nucleon
states. The second relation is obviously valid only in the point form
since in the front one the operators ${\hat U}(l)$ are generally
speaking interaction dependent. The first relation in the point form
is valid too, since, as follows from Eq. (\ref{32}), the operator
${\hat P}$ in $H_0$ is the operator of multiplication by $mG$. We
conclude that $p_i$ has the meaning of the four-momentum of
particle $i$ inside the nucleon only in the point form.

\begin{sloppypar}
 Let $P_i'$ be the initial four momentum of the rest of the target
when the virtual photon is absorbed by particle $i$ and $P_i$ be the
same quantity in the final
state. While the initial state is the nucleon with the four-momentum
$P'$, the final state is the state of free particles with the
four-momentum $P$. If the virtual photon with the four-momentum $q$
is absorbed by particle $i$, then IA for the operator
${\hat J}^{\mu}(0)$ implies that the four-momentum of the spectator
does not change, i.e.
\begin{equation}
P_i'=P_i,\quad P=P'+q
\label{80}
\end{equation}
where the second relation follows from the total four-momentum
conservation. It is easy to see that Eq. (\ref{43}) is a consequence
of Eq. (\ref{80}) in the front form while Eq. (\ref{66}) is a
consequence of Eq. (\ref{80}) in the point form. As shown in
the preceding section, the parton model result $\xi_i'=x$ follows from
Eq. (\ref{43}) while, as shown in this section, the relation between
$\xi_i'$ and $x$ given by Eq. (\ref{70}) follows from Eq. (\ref{66}).
Taking into account the
discussion in the preceding paragraph we again conclude that IA
for the operator ${\hat J}^{\mu}(0)$
has the physical meaning only in the point form, and the derivation
of the relation $\xi_i'=x$ in the parton model is not substantiated.
Our discussion also gives grounds to think that, although the
expression (\ref{73}) for the hadronic tensor is essentially model
dependent, the relation (\ref{70}) is in fact only kinematical since
it does not depend on any specific features of the rest of the target
(in particular on whether $N$ is finite or infinite).
Therefore, if we assume that the final state interaction is not
important in the Bjorken limit, then the relation (\ref{70}) is
rather general, but in the general case $M_i$ entering into the
expression for $M_0$ is the physical (rather than the free) mass
of the rest of the target.
\end{sloppypar}

\section{Deviation from the standard sum rules}
\label{S6}

 If we assume that Eq. (\ref{11}) is valid beyond perturbation
theory (although, as noted in Sect. \ref{S2}, there are serious
reasons to doubt whether this is the case) but no form of the
operators ${\hat O}_i^{\mu_1...\mu_n}$ is prescribed then all
standard results about the $Q^2$ evolution of the structure functions
remain. Indeed in this case the only information about the operators
${\hat O}_i^{\mu_1...\mu_n}$ we need is their tensor
structure since we should correctly parametrize the matrix elements
$\langle P',\chi|{\hat O}_i^{\mu_1\cdots \mu_n}|P',\chi\rangle$.

 However the derivation of sum rules in DIS requires additional
assumptions. It is well-known
that they are derived with different extent of rigor. For example,
the Gottfried and Ellis-Jaffe sum rules \cite{Got,EJ} are essentially
based on model assumptions, the sum rule \cite{Adl} was originally
derived in the framework of current algebra for the time component
of the current operator while the sum rules \cite{Bj1,Bj2,GLS} also
involve the space components. Therefore in the framework of current
algebra the sum rule \cite{Adl} is substantiated in greater extent
than the sum rules \cite{Bj1,Bj2,GLS} (for a detailed discussion see
Refs. \cite{GM,AD,J}). Moreover, there exist field theory models
where ${\hat J}^0({\bf x})$ is free while ${\hat {\bf J}}({\bf x})$
is necessarily interaction dependent
(see e.g. the calculations in scalar QED in Ref. \cite{hep}).
Now the sum rules \cite{Adl,Bj1,Bj2,GLS} are usually considered in the
framework of the OPE and they
have the status of fundamental relations which in fact unambiguously
follow from QCD. However the important assumption in deriving the sum
rules is that the expression for ${\hat O}_V^{\mu}$ coincides with
${\hat J}^{\mu}$, the expression for ${\hat O}_A^{\mu}$ coincides with
the axial current operator ${\hat J}_A^{\mu}$ etc.
(see Eq. (\ref{12})). Our consideration in Sect. \ref{S2} shows that this
assumption has no physical ground. Therefore although (for some reasons)
there may exist sum rules which are satisfied with a good accuracy, the
statement that the sum rules \cite{Adl,Bj1,Bj2,GLS} unambiguously follow
from QCD is not substantiated.

 As already noted, when  $Q^2$ is large, the results of the present
theory agree with the parton model up to anomalous dimensions and
perturbative QCD corrections. The sum rules \cite{Adl,Bj1,Bj2,GLS}
have the property that the corresponding anomalous dimensions are
equal to zero. Therefore, if $Q^2$ is so large that $\alpha_s(Q^2)$
is small, the sum rules \cite{Adl,Bj1,Bj2,GLS} are in agreement
with the parton model.

 The existing experimental data do not make it possible to verify the
sum rules \cite{Adl,Bj2} with a good accuracy, and rather precise
data exist for the sum rules \cite{Bj1,GLS}.

 The data recently obtained by the CCFR collaboration \cite{CCFR} show
that the experimental value of the Gross-Llewellyn Smith sum $S_{GLS}$
is smaller than the value $S_{GLS}=6$ predicted by the parton model.
The analysis of the CCFR data in papers \cite{CCFR,KS,Dor1} shows that
actually $S_{GLS}=4.90\pm 0.16\pm 0.16$ at $Q^2=3\cdot GeV^2$
which is smaller than the value for $S_{GLS}$ which follows from the
present theory, even if the corrections of order $\alpha_s(Q^2)$ and
$\alpha_s(Q^2)^2$ are taken into account.

 The data on the Bjorken sum rule \cite{Bj1} have been analyzed by
several authors (see e.g. Refs.
\cite{Karl,Ellis,Ioffe1,Karl1,RVD,Dor2,Ans,Lampe}).
While some authors argue that the data on the Bjorken sum $S_B$ are in
agreement with the theoretical prediction 0.20, another authors
(see e.g. Ref. \cite{Lampe}) argue that the value consistent with the
data of three experimental groups is 0.15.

 It is important to
note that the data on the sum rules \cite{Bj1,GLS} at small values
of the Bjorken variable $x$ are known only at rather small values of
$Q^2$ and the higher twist corrections are essentially model
dependent. The problem also exists whether the extrapolation of
these data to low $x$ is correct (see the discussion in Refs.
\cite{Karl1,RVD,Dor2,Lampe}). In addition, the problem exists whether
it is possible to extract the neutron structure functions from the
proton and deuteron data \cite{nucl}.

 Let us now consider the sum rules which are not considered fundamental
and one way or another are based on the parton model. The present data
make it possible to unambiguously conclude that the sum rules
\cite{Got,EJ} are not satisfied.  Namely, according to the recent
precise results of the NMC collaboration
\cite{NMC} the experimental value of the integral defining the Gottfried
sum rule \cite{Got} is equal to $0.235\pm 0.026$ instead of 1/3 in the
parton model, and the EMC result \cite{EMC1} for the first moment of the
proton polarized structure function $g_1(x)$ is $\Gamma_p=0.126\pm 0.010
\pm 0.015$ while the Ellis-Jaffe sum rule \cite{EJ} predicts
$\Gamma_p^{EJ}=0.171\pm 0.004$.

 The most impressive results of the parton model are those
concerning the quark contribution to the nucleon momentum and spin.

 The first result (see, for example, the discussions in Ref.
\cite{Close}) says that quarks carry only 46\% of the nucleon
momentum, and this fact is usually considered as one of those which
demonstrates the existence of gluons. In the framework of the OPE the
gluon contribution to the nucleon momentum also can be calculated
(see e.g. Refs. \cite{GW,LopYnd,Ynd}) and the result is in agreement
with the data. However the theoretical formulas are not well
substantiated since corrections to them have the form
$K\alpha_s(Q^2)^{-d}$ where $d>0$ and the coefficient $K$ is not
known \cite{LopYnd,Ynd}.

 The second result known as "the spin crisis" says that the quark
contribution to the nucleon spin is comparable with zero (a detailed
discussion of the spin crisis can be found, for example, in Refs.
\cite{Karl,Karl1,Ellis,Ioffe1,Dor2}).

 Of course, these results are not in direct contradiction with
constituent quark models since the latter are successful only at low
energies. Nevertheless, our experience can be hardly reconciled with
the fact that the role of gluons is so high.

 The above discussion gives grounds to conclude that in the parton
model the values given by the sum rules systematically exceed the
corresponding experimental quantities while the quark contribution
to the nucleon momentum and spin is underestimated.

 The standard sum rules use the fact that in the parton model
$\xi_i'=x$, and the functions $\rho_i(x)$ (see Eqs. (\ref{49}))
satisfy the normalization condition
\begin{equation}
\int_{0}^{1} \rho_i(x) dx =1
\label{81}
\end{equation}
However in our model the structure functions depend on the functions
$\tilde{\rho}_i(x)$ (see Eqs. (\ref{74}-\ref{76})) which do not
satisfy such a condition. The matter is that, as follows from Eqs.
(\ref{71}) and (\ref{72}), the DIS data do not make it possible to
determine the quark distribution at $\xi_i' < \xi_i^{min}$. The
normalization integrals contain the integration over $\xi_i' \in [0,1]$
while the DIS data make it possible to determine some integrals over
$x\in [0,1]$.

 Using Eqs. (\ref{22}), (\ref{26}), (\ref{70}), (\ref{74}) and
changing the integration variable from $x$ to $\xi_i'$ we get
\begin{eqnarray}
&& \tilde{\rho}_i \equiv \int_{0}^{1}\tilde{\rho}_i(x) dx =
 \sum_{\sigma_1,...\sigma_N} \int\nolimits
\frac{d^2{\bf k}_{i\bot}d\rho^F(int_i)}{(2\pi)^3}
\int_{\xi_i^{min}}^{1} \frac{d\xi_i'}{2\xi_i'(1-\xi_i')}\cdot\nonumber\\
&& (\frac{1-x}{1-\xi_i'})^3
|\chi^F({\bf k}_{i\bot},\xi_i,int_i,\sigma_1,...\sigma_N)|^2
[1+\frac{k_i^{'z}}{\omega_i({\bf k}_i')}]
\label{82}
\end{eqnarray}
As argued in the preceding section, this expression is essentially
model dependent but the relation (\ref{70}) is rather general. In
particular $\xi_i^{min}$ can considerably differ from zero.
Therefore, comparing Eq. (\ref{82}) with Eqs. (\ref{49}) and (\ref{81}),
it is natural to expect that $\tilde{\rho}_i<1$.

  Let us consider, for example, the Gottfried sum rule \cite{Got},
according to which the quantity
\begin{equation}
S_G=\int\nolimits [F_{2p}(x)-F_{2n}(x)]\frac{dx}{x}
\label{83}
\end{equation}
is equal to 1/3. Here $F_{2p}(x)$ and $F_{2n}(x)$ are the structure
functions $F_2$ for the proton and neutron respectively. This sum rule
easily follows from Eqs. (\ref{49}), (\ref{51}), (\ref{81}) if we
assume that the neutron wave function can be obtained from the proton
one if one of the $u$ quarks in the proton is replaced by the $d$ quark.
We suppose that particle 1 in the proton is the $u$ quark, particle
1 in the neutron is the $d$ quark and all other particles are the same.
Then, as follows from Eqs. (\ref{75}), (\ref{82}) and (\ref{83}),
$S_G=\tilde{\rho}_1/3$. Therefore, if $\tilde{\rho}_1<1$ then
$S_G<1/3$.

 The Gross-Llewellyn Smith sum rule \cite{GLS} reads
\begin{equation}
S_{GLS}=\int_{0}^{1}[F_3^{\bar{\nu} p}(x)+F_3^{\nu p}(x)]dx=
6[1+(...)]
\label{84}
\end{equation}
where (...) stands for the terms of order $\alpha_s(Q^2)$,
$\alpha_s(Q^2)^2$
etc. It is obvious from Eqs. (\ref{53}) and (\ref{81}) that the
results of the parton model agree with Eq. (\ref{84}) in leading order
in $\alpha_s(Q^2)$. However, our result, which follows from Eqs.
(\ref{76}) and (\ref{82}), is that if, for simplicity, we assume that
the values of $\tilde{\rho}_i$ for the sea quarks and the corresponding
antiquarks are the same, then in leading order in $\alpha_s(Q^2)$
\begin{equation}
S_{GLS}=4\tilde{\rho}_u+2\tilde{\rho}_d
\label{85}
\end{equation}
where $\tilde{\rho}_u$ and $\tilde{\rho}_d$ are the quantities
$\tilde{\rho}_i$ for the valence $u$ and $d$ quarks respectively.
Since it is natural to expect that $\tilde{\rho}_u,\tilde{\rho}_d<1$,
one also might expect that $S_{GLS}<6$.

\begin{sloppypar}
 Analogously, the DIS data alone do not make it possible to determine the
contributions of the $u$, $d$ and $s$ quarks to the nucleon spin, and the
problem of the spin crisis does not arise. Indeed, these contributions
(usually denoted as
$\Delta q=(\Delta u,\Delta d,\Delta s)$) are given by some integrals
over $\xi_i \in [0,1]$. Since the integrals over
$x$ can be transformed to the integrals over
$\xi_i \in [\xi_i^{min},1]$, we see that the DIS data do not make it
possible to determine the contributions of $\xi_i \in [0,\xi_i^{min}]$
to $\Delta q$.
Thus it is natural to expect that the parton model underestimates the
quantities $\Delta q$.
\end{sloppypar}

\begin{sloppypar}
As noted by many authors, the Gottfried sum rule is not a consequence
of (perturbative) QCD. Therefore it is not strange that this sum rule
disagrees with the experimental data, and different explanations of the
disagreement were proposed. Analogously, there exist many papers
devoted to the Ellis-Jaffe sum rule and to the spin crisis. However there
are no approaches explaining why the parton model overestimates the
experimental quantities for all the sum rules and
underestimates the quark contribution to the nucleon momentum and spin.
These facts are qualitatively explained in our model.
\end{sloppypar}

\section{Discussion}
\label{S7}

\begin{sloppypar}
 According to our experience in conventional nuclear and atomic physics,
in processes with high momentum transfer the effect of binding is not
important and the current operator can be taken in IA, i.e. this operator
is the same as for the system of free particles. However this experience
is based on the nonrelativistic quantum mechanics
where only the Hamiltonian is interaction dependent and the other nine
representation generators of the Galilei group are free, while in the
relativistic case at least three representation generators of the
Poincare group are interaction dependent (see Sect. \ref{S1}).
The usual motivation of IA is that if $Q$ is very large then the
process of absorption of the virtual photon is so quick that the struck
quark does not interact with other constituents during this process.
Such a motivation is reasonable in the nonrelativistic case where the
kinetic energies and interaction operators in question are much smaller
than the masses of the constituents. However in the relativistic case
the off-shellness of the struck quark is important and the properties
of the off-shell quark are not the same as of the free one.
\end{sloppypar}

 The parton model is equivalent to IA in the front form of dynamics
(see the discussion in Sect. \ref{S4}). The usual motivation of the
parton model is that, the partons in the IMF are free to the extent
that IA is valid though the partons form the bound state---the nucleon.
It is stated that such a duality is a consequence of asymptotic freedom.
In our opinion, this argument is not convincing since asymptotic freedom
is the argument in favor of using perturbation theory in $\alpha_s(Q^2)$
but the bound state cannot be considered in such a framework.

 In the present theory of DIS the parton model is a consequence of the
factorization theorem \cite{ER,EFP,CSS,Sterman} and the OPE \cite{Wil}.
It is
well-known that the OPE has been proved only in perturbation theory
but many physicists believe that there can be no doubt about the
validity of the OPE beyond that theory. However the consideration in
Sect. \ref{S2} shows the OPE beyond perturbation theory is problematic
and in the general case the factorization theorem does not work.
Therefore the current operator contains a nontrivial nonperturbative
part which contributes to DIS even in the Bjorken limit. Since this
contribution cannot be determined at the present stage of QCD, it is
reasonable to consider models in which the
representation operators of the extended Poincare and the current
operator can be explicitly constructed and it is possible to verify
that they satisfy proper commutation relations.

The important fact (which in our opinion has been overlooked by
physicists working on DIS) is that {\it in the parton model the
conditions (\ref{6}) and (\ref{36}) are not satisfied and therefore
Lorentz invariance, P invariance and T invariance of the current
operator are violated}. This fact has been explained in detail in
Sects. \ref{S2}, \ref{S4} and \ref{S5}. What is the extent of the
violation
of these symmetries in the parton model? The results obtained in the
parton model are the same as in the OPE if the anomalous dimensions are
discarded and only the leading terms of perturbation theory are taken
into account. Since the anomalous dimensions can be determined in the
framework of perturbative QCD, the
difference between the current operators in the OPE and in the parton
model also can be determined in such a framework. Meanwhile the
consideration in Sects. \ref{S4} and \ref{S5} shows that the reason
of the violation of extended Lorentz invariance in the parton model is
the difference between the real mass operator ${\hat M}$ and the free
mass operator $M_0$. This difference is responsible for binding of
quarks and gluons in the nucleon and it cannot be determined in the
framework of perturbative QCD. This observation is the additional
argument that perturbative QCD does not apply to DIS.

 In Sect. \ref{S5} we consider a model in which IA for the operator
${\hat J}^{\mu}(0)$ is compatible with extended Poincare invariance
and current conservation. While the model cannot have any
pretensions to be fundamental, we argue that the relation (\ref{70})
between the light cone momentum fraction $\xi_i'$ and the Bjorken
variable $x$ is in fact kinematical. Therefore DIS experiments
alone cannot determine the $\xi_i'$ distribution of quarks in the
nucleon. This
conclusion also poses the problem whether the parton densities
extracted from the data on DIS at the assumption $\xi_i'=x$ can be
used for describing Drell-Yan processes, hard $p\bar{p}$ collisions
etc.

 If the bound state under consideration can be described
nonrelativistically then the results of the parton model and our one
are the same since in the nonrelativistic approximation
$M_0\approx m\approx m_1+...m_N$. However there is no reason to think
that the nucleon is the nonrelativistic system of quarks and gluons.

 The arguments that nonperturbative effects in the current operator
are important even in the Bjorken limit were given by several
authors (see e.g. Refs. \cite{Misra,GQ,Barn,Prep1,Prep2,Prep2a,Prep3}).
The main difference between our consideration and the consideration
in these references is that we consider in detail the restrictions
imposed on the current operator by its commutation relations with
the representation operators of the extended Poincare group. In the
conventional theory of DIS as well it has never been verified that the
factorization theorem and the OPE beyond perturbation theory are
compatible with these restrictions.

 Although there is no doubt that predictions of perturbative QCD are
in qualitative agreement with experimental data, the extent to which
the agreement is quantitative is not quite clear (we do not discuss
this question in the present paper). Therefore it is very important
to carry out experiments the results of which will undoubtedly show
wether perturbative QCD applies to DIS. In particular it is very
important to test the sum rules \cite{Adl,Bj1,Bj2,GLS}.
However for the unambiguous test they should be extracted
directly from the experimental data at large $Q^2$, not using the
$Q^2$ evolution determined from the OPE or Altarelli-Parisi equations
\cite{AP}. As argued in Ref. \cite{nucl}, it is also not clear whether
the neutron structure functions at small $x$ can be extracted from the
proton and deuteron data even in principle. In this reference we also
argue that the experiment which will show whether the factorization
property takes place, is deuteron DIS at large $Q^2$ and small $x$.

\vskip 1em
\begin{center} {\bf Acknowledgments} \end{center}
\vskip 0.2em

\begin{sloppypar}
 The author is grateful to B.L.G.Bakker, A.Buchmann, F.Coester,
R.van Dantzig, A.E.Dorokhov, L.L.Frankfurt, S.B.Gerasimov, S.Gevorkyan,
I.L.Grach, F.Gross, A.V.Efremov, B.L.Ioffe, A.B.Kaidalov, L.P.Kaptari,
M.Karliner, V.A.Karmanov, N.I.Kochelev, L.A.Kondratyuk,
B.Z.Kopeliovich, S.Kulagin,
E.A.Kuraev, E.Leader, G.I.Lykasov, A.Makhlin, S.V.Mikhailov, P.Mulders,
I.M.Narodetskii, N.N.Nikolaev, V.A.Novikov, E.Pace, G.Salme,
M.G.Sapozhnikov, N.B.Skachkov, S.Simula, O.V.Teryaev, Y.N.Uzikov and
H.J.Weber for valuable discussions and to S.J.Brodsky, S.D.Glazek  and
M.P.Locher for useful remarks. This work was supported by grant No.
96-02-16126a from the Russian Foundation for Fundamental Research.
\end{sloppypar}

\end{document}